\documentclass[longauth]{aa}

\usepackage{graphicx}
\usepackage{natbib}
\usepackage{scalerel}

\usepackage[table]{xcolor}

\usepackage{txfonts}
\usepackage[pdfencoding=auto,psdextra]{hyperref}
\hypersetup{
    colorlinks=true,
    linkcolor=blue,
    filecolor=magenta,      
    urlcolor=blue,
    citecolor=blue
}
\usepackage{orcidlink}
\makeatletter
\renewcommand*\aa@pageof{, page \thepage{} of \pageref*{LastPage}}
\makeatother

\usepackage[utf8]{inputenc}

\usepackage[switch, modulo]{lineno}

\usepackage{xcolor}

\usepackage{euclid}

\begin{document}
%
%

\title{Euclid Quick Data Release (Q1)} \subtitle{The Strong Lensing Discovery Engine D -- Double-source-plane lens candidates}


\newcommand{\tpl}{Teapot Lens}
\newcommand{\cdl}{Cosmic Dartboard}
\newcommand{\gsl}{Galileo’s Lens}
\newcommand{\ssl}{Cosmic Ammonite}

\newcommand{\orcid}[1]{\href{https://orcid.org/#1}{\orcidlink{#1}}}		   
\author{Euclid Collaboration: T.~Li\orcid{0009-0005-5008-0381}\thanks{\email{tian.li@port.ac.uk}}\inst{\ref{aff1}}
\and T.~E.~Collett\orcid{0000-0001-5564-3140}\inst{\ref{aff1}}
\and M.~Walmsley\orcid{0000-0002-6408-4181}\inst{\ref{aff2},\ref{aff3}}
\and N.~E.~P.~Lines\orcid{0009-0004-7751-1914}\inst{\ref{aff1}}
\and K.~Rojas\orcid{0000-0003-1391-6854}\inst{\ref{aff4},\ref{aff1}}
\and J.~W.~Nightingale\orcid{0000-0002-8987-7401}\inst{\ref{aff5}}
\and W.~J.~R.~Enzi\inst{\ref{aff1}}
\and L.~A.~Moustakas\orcid{0000-0003-3030-2360}\inst{\ref{aff6}}
\and C.~Krawczyk\orcid{0000-0001-9233-2341}\inst{\ref{aff1}}
\and R.~Gavazzi\orcid{0000-0002-5540-6935}\inst{\ref{aff7},\ref{aff8}}
\and G.~Despali\orcid{0000-0001-6150-4112}\inst{\ref{aff9},\ref{aff10},\ref{aff11}}
\and P.~Holloway\orcid{0009-0002-8896-6100}\inst{\ref{aff12}}
\and S.~Schuldt\orcid{0000-0003-2497-6334}\inst{\ref{aff13},\ref{aff14}}
\and F.~Courbin\orcid{0000-0003-0758-6510}\inst{\ref{aff15},\ref{aff16}}
\and R.~B.~Metcalf\orcid{0000-0003-3167-2574}\inst{\ref{aff9},\ref{aff10}}
\and D.~J.~Ballard\orcid{0009-0003-3198-7151}\inst{\ref{aff1},\ref{aff17}}
\and A.~Verma\orcid{0000-0002-0730-0781}\inst{\ref{aff12}}
\and B.~Cl\'ement\orcid{0000-0002-7966-3661}\inst{\ref{aff18},\ref{aff19}}
\and H.~Degaudenzi\orcid{0000-0002-5887-6799}\inst{\ref{aff20}}
\and A.~Melo\orcid{0000-0002-6449-3970}\inst{\ref{aff21},\ref{aff22}}
\and J.~A.~Acevedo~Barroso\orcid{0000-0002-9654-1711}\inst{\ref{aff18}}
\and L.~Leuzzi\orcid{0009-0006-4479-7017}\inst{\ref{aff9},\ref{aff10}}
\and A.~Manj\'on-Garc\'ia\orcid{0000-0002-7413-8825}\inst{\ref{aff23}}
\and R.~Pearce-Casey\inst{\ref{aff24}}
\and D.~Sluse\orcid{0000-0001-6116-2095}\inst{\ref{aff25}}
\and C.~Tortora\orcid{0000-0001-7958-6531}\inst{\ref{aff26}}
\and R.~Massey\orcid{0000-0002-6085-3780}\inst{\ref{aff27}}
\and G.~Mahler\orcid{0000-0003-3266-2001}\inst{\ref{aff25},\ref{aff28},\ref{aff27}}
\and A.~More\inst{\ref{aff29},\ref{aff30}}
\and N.~Aghanim\orcid{0000-0002-6688-8992}\inst{\ref{aff31}}
\and B.~Altieri\orcid{0000-0003-3936-0284}\inst{\ref{aff32}}
\and A.~Amara\inst{\ref{aff33}}
\and S.~Andreon\orcid{0000-0002-2041-8784}\inst{\ref{aff34}}
\and N.~Auricchio\orcid{0000-0003-4444-8651}\inst{\ref{aff10}}
\and H.~Aussel\orcid{0000-0002-1371-5705}\inst{\ref{aff35}}
\and C.~Baccigalupi\orcid{0000-0002-8211-1630}\inst{\ref{aff36},\ref{aff37},\ref{aff38},\ref{aff39}}
\and M.~Baldi\orcid{0000-0003-4145-1943}\inst{\ref{aff40},\ref{aff10},\ref{aff11}}
\and A.~Balestra\orcid{0000-0002-6967-261X}\inst{\ref{aff41}}
\and S.~Bardelli\orcid{0000-0002-8900-0298}\inst{\ref{aff10}}
\and P.~Battaglia\orcid{0000-0002-7337-5909}\inst{\ref{aff10}}
\and R.~Bender\orcid{0000-0001-7179-0626}\inst{\ref{aff42},\ref{aff43}}
\and F.~Bernardeau\inst{\ref{aff44},\ref{aff8}}
\and A.~Biviano\orcid{0000-0002-0857-0732}\inst{\ref{aff37},\ref{aff36}}
\and A.~Bonchi\orcid{0000-0002-2667-5482}\inst{\ref{aff45}}
\and E.~Branchini\orcid{0000-0002-0808-6908}\inst{\ref{aff46},\ref{aff47},\ref{aff34}}
\and M.~Brescia\orcid{0000-0001-9506-5680}\inst{\ref{aff48},\ref{aff26}}
\and J.~Brinchmann\orcid{0000-0003-4359-8797}\inst{\ref{aff49},\ref{aff50}}
\and S.~Camera\orcid{0000-0003-3399-3574}\inst{\ref{aff51},\ref{aff52},\ref{aff53}}
\and G.~Ca\~nas-Herrera\orcid{0000-0003-2796-2149}\inst{\ref{aff54},\ref{aff55},\ref{aff56}}
\and V.~Capobianco\orcid{0000-0002-3309-7692}\inst{\ref{aff53}}
\and C.~Carbone\orcid{0000-0003-0125-3563}\inst{\ref{aff14}}
\and V.~F.~Cardone\inst{\ref{aff57},\ref{aff58}}
\and J.~Carretero\orcid{0000-0002-3130-0204}\inst{\ref{aff59},\ref{aff60}}
\and S.~Casas\orcid{0000-0002-4751-5138}\inst{\ref{aff61}}
\and M.~Castellano\orcid{0000-0001-9875-8263}\inst{\ref{aff57}}
\and G.~Castignani\orcid{0000-0001-6831-0687}\inst{\ref{aff10}}
\and S.~Cavuoti\orcid{0000-0002-3787-4196}\inst{\ref{aff26},\ref{aff62}}
\and K.~C.~Chambers\orcid{0000-0001-6965-7789}\inst{\ref{aff63}}
\and A.~Cimatti\inst{\ref{aff64}}
\and C.~Colodro-Conde\inst{\ref{aff65}}
\and G.~Congedo\orcid{0000-0003-2508-0046}\inst{\ref{aff66}}
\and C.~J.~Conselice\orcid{0000-0003-1949-7638}\inst{\ref{aff3}}
\and L.~Conversi\orcid{0000-0002-6710-8476}\inst{\ref{aff67},\ref{aff32}}
\and Y.~Copin\orcid{0000-0002-5317-7518}\inst{\ref{aff68}}
\and H.~M.~Courtois\orcid{0000-0003-0509-1776}\inst{\ref{aff69}}
\and M.~Cropper\orcid{0000-0003-4571-9468}\inst{\ref{aff70}}
\and A.~Da~Silva\orcid{0000-0002-6385-1609}\inst{\ref{aff71},\ref{aff72}}
\and G.~De~Lucia\orcid{0000-0002-6220-9104}\inst{\ref{aff37}}
\and A.~M.~Di~Giorgio\orcid{0000-0002-4767-2360}\inst{\ref{aff73}}
\and C.~Dolding\orcid{0009-0003-7199-6108}\inst{\ref{aff70}}
\and H.~Dole\orcid{0000-0002-9767-3839}\inst{\ref{aff31}}
\and F.~Dubath\orcid{0000-0002-6533-2810}\inst{\ref{aff20}}
\and C.~A.~J.~Duncan\orcid{0009-0003-3573-0791}\inst{\ref{aff3}}
\and X.~Dupac\inst{\ref{aff32}}
\and S.~Escoffier\orcid{0000-0002-2847-7498}\inst{\ref{aff74}}
\and M.~Fabricius\orcid{0000-0002-7025-6058}\inst{\ref{aff42},\ref{aff43}}
\and M.~Farina\orcid{0000-0002-3089-7846}\inst{\ref{aff73}}
\and R.~Farinelli\inst{\ref{aff10}}
\and F.~Faustini\orcid{0000-0001-6274-5145}\inst{\ref{aff45},\ref{aff57}}
\and S.~Ferriol\inst{\ref{aff68}}
\and F.~Finelli\orcid{0000-0002-6694-3269}\inst{\ref{aff10},\ref{aff75}}
\and S.~Fotopoulou\orcid{0000-0002-9686-254X}\inst{\ref{aff76}}
\and M.~Frailis\orcid{0000-0002-7400-2135}\inst{\ref{aff37}}
\and E.~Franceschi\orcid{0000-0002-0585-6591}\inst{\ref{aff10}}
\and S.~Galeotta\orcid{0000-0002-3748-5115}\inst{\ref{aff37}}
\and K.~George\orcid{0000-0002-1734-8455}\inst{\ref{aff43}}
\and W.~Gillard\orcid{0000-0003-4744-9748}\inst{\ref{aff74}}
\and B.~Gillis\orcid{0000-0002-4478-1270}\inst{\ref{aff66}}
\and C.~Giocoli\orcid{0000-0002-9590-7961}\inst{\ref{aff10},\ref{aff11}}
\and P.~G\'omez-Alvarez\orcid{0000-0002-8594-5358}\inst{\ref{aff77},\ref{aff32}}
\and J.~Gracia-Carpio\inst{\ref{aff42}}
\and B.~R.~Granett\orcid{0000-0003-2694-9284}\inst{\ref{aff34}}
\and A.~Grazian\orcid{0000-0002-5688-0663}\inst{\ref{aff41}}
\and F.~Grupp\inst{\ref{aff42},\ref{aff43}}
\and S.~V.~H.~Haugan\orcid{0000-0001-9648-7260}\inst{\ref{aff78}}
\and H.~Hoekstra\orcid{0000-0002-0641-3231}\inst{\ref{aff56}}
\and W.~Holmes\inst{\ref{aff6}}
\and I.~M.~Hook\orcid{0000-0002-2960-978X}\inst{\ref{aff79}}
\and F.~Hormuth\inst{\ref{aff80}}
\and A.~Hornstrup\orcid{0000-0002-3363-0936}\inst{\ref{aff81},\ref{aff82}}
\and P.~Hudelot\inst{\ref{aff8}}
\and K.~Jahnke\orcid{0000-0003-3804-2137}\inst{\ref{aff83}}
\and M.~Jhabvala\inst{\ref{aff84}}
\and B.~Joachimi\orcid{0000-0001-7494-1303}\inst{\ref{aff85}}
\and E.~Keih\"anen\orcid{0000-0003-1804-7715}\inst{\ref{aff86}}
\and S.~Kermiche\orcid{0000-0002-0302-5735}\inst{\ref{aff74}}
\and A.~Kiessling\orcid{0000-0002-2590-1273}\inst{\ref{aff6}}
\and B.~Kubik\orcid{0009-0006-5823-4880}\inst{\ref{aff68}}
\and M.~K\"ummel\orcid{0000-0003-2791-2117}\inst{\ref{aff43}}
\and M.~Kunz\orcid{0000-0002-3052-7394}\inst{\ref{aff87}}
\and H.~Kurki-Suonio\orcid{0000-0002-4618-3063}\inst{\ref{aff88},\ref{aff89}}
\and Q.~Le~Boulc'h\inst{\ref{aff90}}
\and A.~M.~C.~Le~Brun\orcid{0000-0002-0936-4594}\inst{\ref{aff91}}
\and D.~Le~Mignant\orcid{0000-0002-5339-5515}\inst{\ref{aff7}}
\and S.~Ligori\orcid{0000-0003-4172-4606}\inst{\ref{aff53}}
\and P.~B.~Lilje\orcid{0000-0003-4324-7794}\inst{\ref{aff78}}
\and V.~Lindholm\orcid{0000-0003-2317-5471}\inst{\ref{aff88},\ref{aff89}}
\and I.~Lloro\orcid{0000-0001-5966-1434}\inst{\ref{aff92}}
\and G.~Mainetti\orcid{0000-0003-2384-2377}\inst{\ref{aff90}}
\and D.~Maino\inst{\ref{aff13},\ref{aff14},\ref{aff93}}
\and E.~Maiorano\orcid{0000-0003-2593-4355}\inst{\ref{aff10}}
\and O.~Mansutti\orcid{0000-0001-5758-4658}\inst{\ref{aff37}}
\and S.~Marcin\inst{\ref{aff94}}
\and O.~Marggraf\orcid{0000-0001-7242-3852}\inst{\ref{aff95}}
\and M.~Martinelli\orcid{0000-0002-6943-7732}\inst{\ref{aff57},\ref{aff58}}
\and N.~Martinet\orcid{0000-0003-2786-7790}\inst{\ref{aff7}}
\and F.~Marulli\orcid{0000-0002-8850-0303}\inst{\ref{aff9},\ref{aff10},\ref{aff11}}
\and S.~Maurogordato\inst{\ref{aff96}}
\and E.~Medinaceli\orcid{0000-0002-4040-7783}\inst{\ref{aff10}}
\and S.~Mei\orcid{0000-0002-2849-559X}\inst{\ref{aff97},\ref{aff98}}
\and Y.~Mellier\inst{\ref{aff99},\ref{aff8}}
\and M.~Meneghetti\orcid{0000-0003-1225-7084}\inst{\ref{aff10},\ref{aff11}}
\and E.~Merlin\orcid{0000-0001-6870-8900}\inst{\ref{aff57}}
\and G.~Meylan\inst{\ref{aff18}}
\and A.~Mora\orcid{0000-0002-1922-8529}\inst{\ref{aff100}}
\and M.~Moresco\orcid{0000-0002-7616-7136}\inst{\ref{aff9},\ref{aff10}}
\and L.~Moscardini\orcid{0000-0002-3473-6716}\inst{\ref{aff9},\ref{aff10},\ref{aff11}}
\and R.~Nakajima\orcid{0009-0009-1213-7040}\inst{\ref{aff95}}
\and C.~Neissner\orcid{0000-0001-8524-4968}\inst{\ref{aff101},\ref{aff60}}
\and R.~C.~Nichol\orcid{0000-0003-0939-6518}\inst{\ref{aff33}}
\and S.-M.~Niemi\inst{\ref{aff54}}
\and C.~Padilla\orcid{0000-0001-7951-0166}\inst{\ref{aff101}}
\and S.~Paltani\orcid{0000-0002-8108-9179}\inst{\ref{aff20}}
\and F.~Pasian\orcid{0000-0002-4869-3227}\inst{\ref{aff37}}
\and K.~Pedersen\inst{\ref{aff102}}
\and W.~J.~Percival\orcid{0000-0002-0644-5727}\inst{\ref{aff103},\ref{aff104},\ref{aff105}}
\and V.~Pettorino\inst{\ref{aff54}}
\and S.~Pires\orcid{0000-0002-0249-2104}\inst{\ref{aff35}}
\and G.~Polenta\orcid{0000-0003-4067-9196}\inst{\ref{aff45}}
\and M.~Poncet\inst{\ref{aff106}}
\and L.~A.~Popa\inst{\ref{aff107}}
\and L.~Pozzetti\orcid{0000-0001-7085-0412}\inst{\ref{aff10}}
\and F.~Raison\orcid{0000-0002-7819-6918}\inst{\ref{aff42}}
\and R.~Rebolo\orcid{0000-0003-3767-7085}\inst{\ref{aff65},\ref{aff108},\ref{aff109}}
\and A.~Renzi\orcid{0000-0001-9856-1970}\inst{\ref{aff110},\ref{aff111}}
\and J.~Rhodes\orcid{0000-0002-4485-8549}\inst{\ref{aff6}}
\and G.~Riccio\inst{\ref{aff26}}
\and E.~Romelli\orcid{0000-0003-3069-9222}\inst{\ref{aff37}}
\and M.~Roncarelli\orcid{0000-0001-9587-7822}\inst{\ref{aff10}}
\and R.~Saglia\orcid{0000-0003-0378-7032}\inst{\ref{aff43},\ref{aff42}}
\and Z.~Sakr\orcid{0000-0002-4823-3757}\inst{\ref{aff112},\ref{aff113},\ref{aff114}}
\and D.~Sapone\orcid{0000-0001-7089-4503}\inst{\ref{aff115}}
\and B.~Sartoris\orcid{0000-0003-1337-5269}\inst{\ref{aff43},\ref{aff37}}
\and J.~A.~Schewtschenko\orcid{0000-0002-4913-6393}\inst{\ref{aff66}}
\and M.~Schirmer\orcid{0000-0003-2568-9994}\inst{\ref{aff83}}
\and P.~Schneider\orcid{0000-0001-8561-2679}\inst{\ref{aff95}}
\and T.~Schrabback\orcid{0000-0002-6987-7834}\inst{\ref{aff116}}
\and A.~Secroun\orcid{0000-0003-0505-3710}\inst{\ref{aff74}}
\and G.~Seidel\orcid{0000-0003-2907-353X}\inst{\ref{aff83}}
\and M.~Seiffert\orcid{0000-0002-7536-9393}\inst{\ref{aff6}}
\and S.~Serrano\orcid{0000-0002-0211-2861}\inst{\ref{aff117},\ref{aff118},\ref{aff119}}
\and P.~Simon\inst{\ref{aff95}}
\and C.~Sirignano\orcid{0000-0002-0995-7146}\inst{\ref{aff110},\ref{aff111}}
\and G.~Sirri\orcid{0000-0003-2626-2853}\inst{\ref{aff11}}
\and A.~Spurio~Mancini\orcid{0000-0001-5698-0990}\inst{\ref{aff120}}
\and L.~Stanco\orcid{0000-0002-9706-5104}\inst{\ref{aff111}}
\and J.~Steinwagner\orcid{0000-0001-7443-1047}\inst{\ref{aff42}}
\and P.~Tallada-Cresp\'{i}\orcid{0000-0002-1336-8328}\inst{\ref{aff59},\ref{aff60}}
\and A.~N.~Taylor\inst{\ref{aff66}}
\and I.~Tereno\inst{\ref{aff71},\ref{aff121}}
\and N.~Tessore\orcid{0000-0002-9696-7931}\inst{\ref{aff85}}
\and S.~Toft\orcid{0000-0003-3631-7176}\inst{\ref{aff122},\ref{aff123}}
\and R.~Toledo-Moreo\orcid{0000-0002-2997-4859}\inst{\ref{aff124}}
\and F.~Torradeflot\orcid{0000-0003-1160-1517}\inst{\ref{aff60},\ref{aff59}}
\and I.~Tutusaus\orcid{0000-0002-3199-0399}\inst{\ref{aff113}}
\and E.~A.~Valentijn\inst{\ref{aff125}}
\and L.~Valenziano\orcid{0000-0002-1170-0104}\inst{\ref{aff10},\ref{aff75}}
\and J.~Valiviita\orcid{0000-0001-6225-3693}\inst{\ref{aff88},\ref{aff89}}
\and T.~Vassallo\orcid{0000-0001-6512-6358}\inst{\ref{aff43},\ref{aff37}}
\and G.~Verdoes~Kleijn\orcid{0000-0001-5803-2580}\inst{\ref{aff125}}
\and A.~Veropalumbo\orcid{0000-0003-2387-1194}\inst{\ref{aff34},\ref{aff47},\ref{aff46}}
\and Y.~Wang\orcid{0000-0002-4749-2984}\inst{\ref{aff126}}
\and J.~Weller\orcid{0000-0002-8282-2010}\inst{\ref{aff43},\ref{aff42}}
\and A.~Zacchei\orcid{0000-0003-0396-1192}\inst{\ref{aff37},\ref{aff36}}
\and G.~Zamorani\orcid{0000-0002-2318-301X}\inst{\ref{aff10}}
\and F.~M.~Zerbi\inst{\ref{aff34}}
\and E.~Zucca\orcid{0000-0002-5845-8132}\inst{\ref{aff10}}
\and V.~Allevato\orcid{0000-0001-7232-5152}\inst{\ref{aff26}}
\and M.~Ballardini\orcid{0000-0003-4481-3559}\inst{\ref{aff127},\ref{aff128},\ref{aff10}}
\and M.~Bolzonella\orcid{0000-0003-3278-4607}\inst{\ref{aff10}}
\and E.~Bozzo\orcid{0000-0002-8201-1525}\inst{\ref{aff20}}
\and C.~Burigana\orcid{0000-0002-3005-5796}\inst{\ref{aff129},\ref{aff75}}
\and R.~Cabanac\orcid{0000-0001-6679-2600}\inst{\ref{aff113}}
\and A.~Cappi\inst{\ref{aff10},\ref{aff96}}
\and D.~Di~Ferdinando\inst{\ref{aff11}}
\and J.~A.~Escartin~Vigo\inst{\ref{aff42}}
\and L.~Gabarra\orcid{0000-0002-8486-8856}\inst{\ref{aff12}}
\and M.~Huertas-Company\orcid{0000-0002-1416-8483}\inst{\ref{aff65},\ref{aff130},\ref{aff131},\ref{aff132}}
\and J.~Mart\'{i}n-Fleitas\orcid{0000-0002-8594-569X}\inst{\ref{aff100}}
\and S.~Matthew\orcid{0000-0001-8448-1697}\inst{\ref{aff66}}
\and N.~Mauri\orcid{0000-0001-8196-1548}\inst{\ref{aff64},\ref{aff11}}
\and A.~Pezzotta\orcid{0000-0003-0726-2268}\inst{\ref{aff133},\ref{aff42}}
\and M.~P\"ontinen\orcid{0000-0001-5442-2530}\inst{\ref{aff88}}
\and C.~Porciani\orcid{0000-0002-7797-2508}\inst{\ref{aff95}}
\and I.~Risso\orcid{0000-0003-2525-7761}\inst{\ref{aff134}}
\and V.~Scottez\inst{\ref{aff99},\ref{aff135}}
\and M.~Sereno\orcid{0000-0003-0302-0325}\inst{\ref{aff10},\ref{aff11}}
\and M.~Tenti\orcid{0000-0002-4254-5901}\inst{\ref{aff11}}
\and M.~Viel\orcid{0000-0002-2642-5707}\inst{\ref{aff36},\ref{aff37},\ref{aff39},\ref{aff38},\ref{aff136}}
\and M.~Wiesmann\orcid{0009-0000-8199-5860}\inst{\ref{aff78}}
\and Y.~Akrami\orcid{0000-0002-2407-7956}\inst{\ref{aff137},\ref{aff138}}
\and S.~Alvi\orcid{0000-0001-5779-8568}\inst{\ref{aff127}}
\and I.~T.~Andika\orcid{0000-0001-6102-9526}\inst{\ref{aff22},\ref{aff21}}
\and S.~Anselmi\orcid{0000-0002-3579-9583}\inst{\ref{aff111},\ref{aff110},\ref{aff139}}
\and M.~Archidiacono\orcid{0000-0003-4952-9012}\inst{\ref{aff13},\ref{aff93}}
\and F.~Atrio-Barandela\orcid{0000-0002-2130-2513}\inst{\ref{aff140}}
\and K.~Benson\inst{\ref{aff70}}
\and P.~Bergamini\orcid{0000-0003-1383-9414}\inst{\ref{aff13},\ref{aff10}}
\and D.~Bertacca\orcid{0000-0002-2490-7139}\inst{\ref{aff110},\ref{aff41},\ref{aff111}}
\and M.~Bethermin\orcid{0000-0002-3915-2015}\inst{\ref{aff141}}
\and A.~Blanchard\orcid{0000-0001-8555-9003}\inst{\ref{aff113}}
\and L.~Blot\orcid{0000-0002-9622-7167}\inst{\ref{aff142},\ref{aff139}}
\and M.~L.~Brown\orcid{0000-0002-0370-8077}\inst{\ref{aff3}}
\and S.~Bruton\orcid{0000-0002-6503-5218}\inst{\ref{aff143}}
\and A.~Calabro\orcid{0000-0003-2536-1614}\inst{\ref{aff57}}
\and F.~Caro\inst{\ref{aff57}}
\and C.~S.~Carvalho\inst{\ref{aff121}}
\and T.~Castro\orcid{0000-0002-6292-3228}\inst{\ref{aff37},\ref{aff38},\ref{aff36},\ref{aff136}}
\and F.~Cogato\orcid{0000-0003-4632-6113}\inst{\ref{aff9},\ref{aff10}}
\and A.~R.~Cooray\orcid{0000-0002-3892-0190}\inst{\ref{aff144}}
\and O.~Cucciati\orcid{0000-0002-9336-7551}\inst{\ref{aff10}}
\and S.~Davini\orcid{0000-0003-3269-1718}\inst{\ref{aff47}}
\and F.~De~Paolis\orcid{0000-0001-6460-7563}\inst{\ref{aff145},\ref{aff146},\ref{aff147}}
\and G.~Desprez\orcid{0000-0001-8325-1742}\inst{\ref{aff125}}
\and A.~D\'iaz-S\'anchez\orcid{0000-0003-0748-4768}\inst{\ref{aff23}}
\and J.~J.~Diaz\inst{\ref{aff130}}
\and S.~Di~Domizio\orcid{0000-0003-2863-5895}\inst{\ref{aff46},\ref{aff47}}
\and J.~M.~Diego\orcid{0000-0001-9065-3926}\inst{\ref{aff148}}
\and P.-A.~Duc\orcid{0000-0003-3343-6284}\inst{\ref{aff141}}
\and A.~Enia\orcid{0000-0002-0200-2857}\inst{\ref{aff40},\ref{aff10}}
\and Y.~Fang\inst{\ref{aff43}}
\and A.~G.~Ferrari\orcid{0009-0005-5266-4110}\inst{\ref{aff11}}
\and P.~G.~Ferreira\orcid{0000-0002-3021-2851}\inst{\ref{aff12}}
\and A.~Finoguenov\orcid{0000-0002-4606-5403}\inst{\ref{aff88}}
\and A.~Fontana\orcid{0000-0003-3820-2823}\inst{\ref{aff57}}
\and A.~Franco\orcid{0000-0002-4761-366X}\inst{\ref{aff146},\ref{aff145},\ref{aff147}}
\and K.~Ganga\orcid{0000-0001-8159-8208}\inst{\ref{aff97}}
\and J.~Garc\'ia-Bellido\orcid{0000-0002-9370-8360}\inst{\ref{aff137}}
\and T.~Gasparetto\orcid{0000-0002-7913-4866}\inst{\ref{aff37}}
\and V.~Gautard\inst{\ref{aff149}}
\and E.~Gaztanaga\orcid{0000-0001-9632-0815}\inst{\ref{aff119},\ref{aff117},\ref{aff1}}
\and F.~Giacomini\orcid{0000-0002-3129-2814}\inst{\ref{aff11}}
\and F.~Gianotti\orcid{0000-0003-4666-119X}\inst{\ref{aff10}}
\and G.~Gozaliasl\orcid{0000-0002-0236-919X}\inst{\ref{aff150},\ref{aff88}}
\and M.~Guidi\orcid{0000-0001-9408-1101}\inst{\ref{aff40},\ref{aff10}}
\and C.~M.~Gutierrez\orcid{0000-0001-7854-783X}\inst{\ref{aff151}}
\and A.~Hall\orcid{0000-0002-3139-8651}\inst{\ref{aff66}}
\and W.~G.~Hartley\inst{\ref{aff20}}
\and C.~Hern\'andez-Monteagudo\orcid{0000-0001-5471-9166}\inst{\ref{aff109},\ref{aff65}}
\and H.~Hildebrandt\orcid{0000-0002-9814-3338}\inst{\ref{aff152}}
\and J.~Hjorth\orcid{0000-0002-4571-2306}\inst{\ref{aff102}}
\and M.~Jauzac\orcid{0000-0003-1974-8732}\inst{\ref{aff28},\ref{aff27},\ref{aff153},\ref{aff154}}
\and J.~J.~E.~Kajava\orcid{0000-0002-3010-8333}\inst{\ref{aff155},\ref{aff156}}
\and Y.~Kang\orcid{0009-0000-8588-7250}\inst{\ref{aff20}}
\and V.~Kansal\orcid{0000-0002-4008-6078}\inst{\ref{aff157},\ref{aff158}}
\and D.~Karagiannis\orcid{0000-0002-4927-0816}\inst{\ref{aff127},\ref{aff159}}
\and K.~Kiiveri\inst{\ref{aff86}}
\and C.~C.~Kirkpatrick\inst{\ref{aff86}}
\and S.~Kruk\orcid{0000-0001-8010-8879}\inst{\ref{aff32}}
\and J.~Le~Graet\orcid{0000-0001-6523-7971}\inst{\ref{aff74}}
\and L.~Legrand\orcid{0000-0003-0610-5252}\inst{\ref{aff160},\ref{aff161}}
\and M.~Lembo\orcid{0000-0002-5271-5070}\inst{\ref{aff127},\ref{aff128}}
\and F.~Lepori\orcid{0009-0000-5061-7138}\inst{\ref{aff162}}
\and G.~Leroy\orcid{0009-0004-2523-4425}\inst{\ref{aff28},\ref{aff27}}
\and G.~F.~Lesci\orcid{0000-0002-4607-2830}\inst{\ref{aff9},\ref{aff10}}
\and J.~Lesgourgues\orcid{0000-0001-7627-353X}\inst{\ref{aff61}}
\and T.~I.~Liaudat\orcid{0000-0002-9104-314X}\inst{\ref{aff163}}
\and A.~Loureiro\orcid{0000-0002-4371-0876}\inst{\ref{aff164},\ref{aff165}}
\and J.~Macias-Perez\orcid{0000-0002-5385-2763}\inst{\ref{aff166}}
\and G.~Maggio\orcid{0000-0003-4020-4836}\inst{\ref{aff37}}
\and M.~Magliocchetti\orcid{0000-0001-9158-4838}\inst{\ref{aff73}}
\and F.~Mannucci\orcid{0000-0002-4803-2381}\inst{\ref{aff167}}
\and R.~Maoli\orcid{0000-0002-6065-3025}\inst{\ref{aff168},\ref{aff57}}
\and C.~J.~A.~P.~Martins\orcid{0000-0002-4886-9261}\inst{\ref{aff169},\ref{aff49}}
\and L.~Maurin\orcid{0000-0002-8406-0857}\inst{\ref{aff31}}
\and M.~Migliaccio\inst{\ref{aff170},\ref{aff171}}
\and M.~Miluzio\inst{\ref{aff32},\ref{aff172}}
\and P.~Monaco\orcid{0000-0003-2083-7564}\inst{\ref{aff173},\ref{aff37},\ref{aff38},\ref{aff36}}
\and C.~Moretti\orcid{0000-0003-3314-8936}\inst{\ref{aff39},\ref{aff136},\ref{aff37},\ref{aff36},\ref{aff38}}
\and G.~Morgante\inst{\ref{aff10}}
\and S.~Nadathur\orcid{0000-0001-9070-3102}\inst{\ref{aff1}}
\and K.~Naidoo\orcid{0000-0002-9182-1802}\inst{\ref{aff1}}
\and A.~Navarro-Alsina\orcid{0000-0002-3173-2592}\inst{\ref{aff95}}
\and S.~Nesseris\orcid{0000-0002-0567-0324}\inst{\ref{aff137}}
\and F.~Passalacqua\orcid{0000-0002-8606-4093}\inst{\ref{aff110},\ref{aff111}}
\and K.~Paterson\orcid{0000-0001-8340-3486}\inst{\ref{aff83}}
\and L.~Patrizii\inst{\ref{aff11}}
\and A.~Pisani\orcid{0000-0002-6146-4437}\inst{\ref{aff74},\ref{aff174}}
\and D.~Potter\orcid{0000-0002-0757-5195}\inst{\ref{aff162}}
\and S.~Quai\orcid{0000-0002-0449-8163}\inst{\ref{aff9},\ref{aff10}}
\and M.~Radovich\orcid{0000-0002-3585-866X}\inst{\ref{aff41}}
\and P.-F.~Rocci\inst{\ref{aff31}}
\and S.~Sacquegna\orcid{0000-0002-8433-6630}\inst{\ref{aff145},\ref{aff146},\ref{aff147}}
\and M.~Sahl\'en\orcid{0000-0003-0973-4804}\inst{\ref{aff175}}
\and D.~B.~Sanders\orcid{0000-0002-1233-9998}\inst{\ref{aff63}}
\and E.~Sarpa\orcid{0000-0002-1256-655X}\inst{\ref{aff39},\ref{aff136},\ref{aff38}}
\and C.~Scarlata\orcid{0000-0002-9136-8876}\inst{\ref{aff176}}
\and J.~Schaye\orcid{0000-0002-0668-5560}\inst{\ref{aff56}}
\and A.~Schneider\orcid{0000-0001-7055-8104}\inst{\ref{aff162}}
\and D.~Sciotti\orcid{0009-0008-4519-2620}\inst{\ref{aff57},\ref{aff58}}
\and E.~Sellentin\inst{\ref{aff177},\ref{aff56}}
\and L.~C.~Smith\orcid{0000-0002-3259-2771}\inst{\ref{aff178}}
\and K.~Tanidis\orcid{0000-0001-9843-5130}\inst{\ref{aff12}}
\and G.~Testera\inst{\ref{aff47}}
\and R.~Teyssier\orcid{0000-0001-7689-0933}\inst{\ref{aff174}}
\and S.~Tosi\orcid{0000-0002-7275-9193}\inst{\ref{aff46},\ref{aff47},\ref{aff34}}
\and A.~Troja\orcid{0000-0003-0239-4595}\inst{\ref{aff110},\ref{aff111}}
\and M.~Tucci\inst{\ref{aff20}}
\and C.~Valieri\inst{\ref{aff11}}
\and A.~Venhola\orcid{0000-0001-6071-4564}\inst{\ref{aff179}}
\and D.~Vergani\orcid{0000-0003-0898-2216}\inst{\ref{aff10}}
\and G.~Vernardos\orcid{0000-0001-8554-7248}\inst{\ref{aff180},\ref{aff181}}
\and G.~Verza\orcid{0000-0002-1886-8348}\inst{\ref{aff182}}
\and P.~Vielzeuf\orcid{0000-0003-2035-9339}\inst{\ref{aff74}}
\and N.~A.~Walton\orcid{0000-0003-3983-8778}\inst{\ref{aff178}}
\and J.~Wilde\orcid{0000-0002-4460-7379}\inst{\ref{aff15}}
\and D.~Scott\orcid{0000-0002-6878-9840}\inst{\ref{aff183}}}
										   
\institute{Institute of Cosmology and Gravitation, University of Portsmouth, Portsmouth PO1 3FX, UK\label{aff1}
\and
David A. Dunlap Department of Astronomy \& Astrophysics, University of Toronto, 50 St George Street, Toronto, Ontario M5S 3H4, Canada\label{aff2}
\and
Jodrell Bank Centre for Astrophysics, Department of Physics and Astronomy, University of Manchester, Oxford Road, Manchester M13 9PL, UK\label{aff3}
\and
University of Applied Sciences and Arts of Northwestern Switzerland, School of Engineering, 5210 Windisch, Switzerland\label{aff4}
\and
School of Mathematics, Statistics and Physics, Newcastle University, Herschel Building, Newcastle-upon-Tyne, NE1 7RU, UK\label{aff5}
\and
Jet Propulsion Laboratory, California Institute of Technology, 4800 Oak Grove Drive, Pasadena, CA, 91109, USA\label{aff6}
\and
Aix-Marseille Universit\'e, CNRS, CNES, LAM, Marseille, France\label{aff7}
\and
Institut d'Astrophysique de Paris, UMR 7095, CNRS, and Sorbonne Universit\'e, 98 bis boulevard Arago, 75014 Paris, France\label{aff8}
\and
Dipartimento di Fisica e Astronomia "Augusto Righi" - Alma Mater Studiorum Universit\`a di Bologna, via Piero Gobetti 93/2, 40129 Bologna, Italy\label{aff9}
\and
INAF-Osservatorio di Astrofisica e Scienza dello Spazio di Bologna, Via Piero Gobetti 93/3, 40129 Bologna, Italy\label{aff10}
\and
INFN-Sezione di Bologna, Viale Berti Pichat 6/2, 40127 Bologna, Italy\label{aff11}
\and
Department of Physics, Oxford University, Keble Road, Oxford OX1 3RH, UK\label{aff12}
\and
Dipartimento di Fisica "Aldo Pontremoli", Universit\`a degli Studi di Milano, Via Celoria 16, 20133 Milano, Italy\label{aff13}
\and
INAF-IASF Milano, Via Alfonso Corti 12, 20133 Milano, Italy\label{aff14}
\and
Institut de Ci\`{e}ncies del Cosmos (ICCUB), Universitat de Barcelona (IEEC-UB), Mart\'{i} i Franqu\`{e}s 1, 08028 Barcelona, Spain\label{aff15}
\and
Instituci\'o Catalana de Recerca i Estudis Avan\c{c}ats (ICREA), Passeig de Llu\'{\i}s Companys 23, 08010 Barcelona, Spain\label{aff16}
\and
Sydney Institute for Astronomy, School of Physics, University of Sydney, NSW 2006, Australia\label{aff17}
\and
Institute of Physics, Laboratory of Astrophysics, Ecole Polytechnique F\'ed\'erale de Lausanne (EPFL), Observatoire de Sauverny, 1290 Versoix, Switzerland\label{aff18}
\and
SCITAS, Ecole Polytechnique F\'ed\'erale de Lausanne (EPFL), 1015 Lausanne, Switzerland\label{aff19}
\and
Department of Astronomy, University of Geneva, ch. d'Ecogia 16, 1290 Versoix, Switzerland\label{aff20}
\and
Max-Planck-Institut f\"ur Astrophysik, Karl-Schwarzschild-Str.~1, 85748 Garching, Germany\label{aff21}
\and
Technical University of Munich, TUM School of Natural Sciences, Physics Department, James-Franck-Str.~1, 85748 Garching, Germany\label{aff22}
\and
Departamento F\'isica Aplicada, Universidad Polit\'ecnica de Cartagena, Campus Muralla del Mar, 30202 Cartagena, Murcia, Spain\label{aff23}
\and
School of Physical Sciences, The Open University, Milton Keynes, MK7 6AA, UK\label{aff24}
\and
STAR Institute, University of Li{\`e}ge, Quartier Agora, All\'ee du six Ao\^ut 19c, 4000 Li\`ege, Belgium\label{aff25}
\and
INAF-Osservatorio Astronomico di Capodimonte, Via Moiariello 16, 80131 Napoli, Italy\label{aff26}
\and
Department of Physics, Institute for Computational Cosmology, Durham University, South Road, Durham, DH1 3LE, UK\label{aff27}
\and
Department of Physics, Centre for Extragalactic Astronomy, Durham University, South Road, Durham, DH1 3LE, UK\label{aff28}
\and
The Inter-University Centre for Astronomy and Astrophysics, Post Bag 4, Ganeshkhind, Pune 411007, India\label{aff29}
\and
Kavli Institute for the Physics and Mathematics of the Universe (WPI), University of Tokyo, Kashiwa, Chiba 277-8583, Japan\label{aff30}
\and
Universit\'e Paris-Saclay, CNRS, Institut d'astrophysique spatiale, 91405, Orsay, France\label{aff31}
\and
ESAC/ESA, Camino Bajo del Castillo, s/n., Urb. Villafranca del Castillo, 28692 Villanueva de la Ca\~nada, Madrid, Spain\label{aff32}
\and
School of Mathematics and Physics, University of Surrey, Guildford, Surrey, GU2 7XH, UK\label{aff33}
\and
INAF-Osservatorio Astronomico di Brera, Via Brera 28, 20122 Milano, Italy\label{aff34}
\and
Universit\'e Paris-Saclay, Universit\'e Paris Cit\'e, CEA, CNRS, AIM, 91191, Gif-sur-Yvette, France\label{aff35}
\and
IFPU, Institute for Fundamental Physics of the Universe, via Beirut 2, 34151 Trieste, Italy\label{aff36}
\and
INAF-Osservatorio Astronomico di Trieste, Via G. B. Tiepolo 11, 34143 Trieste, Italy\label{aff37}
\and
INFN, Sezione di Trieste, Via Valerio 2, 34127 Trieste TS, Italy\label{aff38}
\and
SISSA, International School for Advanced Studies, Via Bonomea 265, 34136 Trieste TS, Italy\label{aff39}
\and
Dipartimento di Fisica e Astronomia, Universit\`a di Bologna, Via Gobetti 93/2, 40129 Bologna, Italy\label{aff40}
\and
INAF-Osservatorio Astronomico di Padova, Via dell'Osservatorio 5, 35122 Padova, Italy\label{aff41}
\and
Max Planck Institute for Extraterrestrial Physics, Giessenbachstr. 1, 85748 Garching, Germany\label{aff42}
\and
Universit\"ats-Sternwarte M\"unchen, Fakult\"at f\"ur Physik, Ludwig-Maximilians-Universit\"at M\"unchen, Scheinerstrasse 1, 81679 M\"unchen, Germany\label{aff43}
\and
Institut de Physique Th\'eorique, CEA, CNRS, Universit\'e Paris-Saclay 91191 Gif-sur-Yvette Cedex, France\label{aff44}
\and
Space Science Data Center, Italian Space Agency, via del Politecnico snc, 00133 Roma, Italy\label{aff45}
\and
Dipartimento di Fisica, Universit\`a di Genova, Via Dodecaneso 33, 16146, Genova, Italy\label{aff46}
\and
INFN-Sezione di Genova, Via Dodecaneso 33, 16146, Genova, Italy\label{aff47}
\and
Department of Physics "E. Pancini", University Federico II, Via Cinthia 6, 80126, Napoli, Italy\label{aff48}
\and
Instituto de Astrof\'isica e Ci\^encias do Espa\c{c}o, Universidade do Porto, CAUP, Rua das Estrelas, PT4150-762 Porto, Portugal\label{aff49}
\and
Faculdade de Ci\^encias da Universidade do Porto, Rua do Campo de Alegre, 4150-007 Porto, Portugal\label{aff50}
\and
Dipartimento di Fisica, Universit\`a degli Studi di Torino, Via P. Giuria 1, 10125 Torino, Italy\label{aff51}
\and
INFN-Sezione di Torino, Via P. Giuria 1, 10125 Torino, Italy\label{aff52}
\and
INAF-Osservatorio Astrofisico di Torino, Via Osservatorio 20, 10025 Pino Torinese (TO), Italy\label{aff53}
\and
European Space Agency/ESTEC, Keplerlaan 1, 2201 AZ Noordwijk, The Netherlands\label{aff54}
\and
Institute Lorentz, Leiden University, Niels Bohrweg 2, 2333 CA Leiden, The Netherlands\label{aff55}
\and
Leiden Observatory, Leiden University, Einsteinweg 55, 2333 CC Leiden, The Netherlands\label{aff56}
\and
INAF-Osservatorio Astronomico di Roma, Via Frascati 33, 00078 Monteporzio Catone, Italy\label{aff57}
\and
INFN-Sezione di Roma, Piazzale Aldo Moro, 2 - c/o Dipartimento di Fisica, Edificio G. Marconi, 00185 Roma, Italy\label{aff58}
\and
Centro de Investigaciones Energ\'eticas, Medioambientales y Tecnol\'ogicas (CIEMAT), Avenida Complutense 40, 28040 Madrid, Spain\label{aff59}
\and
Port d'Informaci\'{o} Cient\'{i}fica, Campus UAB, C. Albareda s/n, 08193 Bellaterra (Barcelona), Spain\label{aff60}
\and
Institute for Theoretical Particle Physics and Cosmology (TTK), RWTH Aachen University, 52056 Aachen, Germany\label{aff61}
\and
INFN section of Naples, Via Cinthia 6, 80126, Napoli, Italy\label{aff62}
\and
Institute for Astronomy, University of Hawaii, 2680 Woodlawn Drive, Honolulu, HI 96822, USA\label{aff63}
\and
Dipartimento di Fisica e Astronomia "Augusto Righi" - Alma Mater Studiorum Universit\`a di Bologna, Viale Berti Pichat 6/2, 40127 Bologna, Italy\label{aff64}
\and
Instituto de Astrof\'{\i}sica de Canarias, V\'{\i}a L\'actea, 38205 La Laguna, Tenerife, Spain\label{aff65}
\and
Institute for Astronomy, University of Edinburgh, Royal Observatory, Blackford Hill, Edinburgh EH9 3HJ, UK\label{aff66}
\and
European Space Agency/ESRIN, Largo Galileo Galilei 1, 00044 Frascati, Roma, Italy\label{aff67}
\and
Universit\'e Claude Bernard Lyon 1, CNRS/IN2P3, IP2I Lyon, UMR 5822, Villeurbanne, F-69100, France\label{aff68}
\and
UCB Lyon 1, CNRS/IN2P3, IUF, IP2I Lyon, 4 rue Enrico Fermi, 69622 Villeurbanne, France\label{aff69}
\and
Mullard Space Science Laboratory, University College London, Holmbury St Mary, Dorking, Surrey RH5 6NT, UK\label{aff70}
\and
Departamento de F\'isica, Faculdade de Ci\^encias, Universidade de Lisboa, Edif\'icio C8, Campo Grande, PT1749-016 Lisboa, Portugal\label{aff71}
\and
Instituto de Astrof\'isica e Ci\^encias do Espa\c{c}o, Faculdade de Ci\^encias, Universidade de Lisboa, Campo Grande, 1749-016 Lisboa, Portugal\label{aff72}
\and
INAF-Istituto di Astrofisica e Planetologia Spaziali, via del Fosso del Cavaliere, 100, 00100 Roma, Italy\label{aff73}
\and
Aix-Marseille Universit\'e, CNRS/IN2P3, CPPM, Marseille, France\label{aff74}
\and
INFN-Bologna, Via Irnerio 46, 40126 Bologna, Italy\label{aff75}
\and
School of Physics, HH Wills Physics Laboratory, University of Bristol, Tyndall Avenue, Bristol, BS8 1TL, UK\label{aff76}
\and
FRACTAL S.L.N.E., calle Tulip\'an 2, Portal 13 1A, 28231, Las Rozas de Madrid, Spain\label{aff77}
\and
Institute of Theoretical Astrophysics, University of Oslo, P.O. Box 1029 Blindern, 0315 Oslo, Norway\label{aff78}
\and
Department of Physics, Lancaster University, Lancaster, LA1 4YB, UK\label{aff79}
\and
Felix Hormuth Engineering, Goethestr. 17, 69181 Leimen, Germany\label{aff80}
\and
Technical University of Denmark, Elektrovej 327, 2800 Kgs. Lyngby, Denmark\label{aff81}
\and
Cosmic Dawn Center (DAWN), Denmark\label{aff82}
\and
Max-Planck-Institut f\"ur Astronomie, K\"onigstuhl 17, 69117 Heidelberg, Germany\label{aff83}
\and
NASA Goddard Space Flight Center, Greenbelt, MD 20771, USA\label{aff84}
\and
Department of Physics and Astronomy, University College London, Gower Street, London WC1E 6BT, UK\label{aff85}
\and
Department of Physics and Helsinki Institute of Physics, Gustaf H\"allstr\"omin katu 2, 00014 University of Helsinki, Finland\label{aff86}
\and
Universit\'e de Gen\`eve, D\'epartement de Physique Th\'eorique and Centre for Astroparticle Physics, 24 quai Ernest-Ansermet, CH-1211 Gen\`eve 4, Switzerland\label{aff87}
\and
Department of Physics, P.O. Box 64, 00014 University of Helsinki, Finland\label{aff88}
\and
Helsinki Institute of Physics, Gustaf H{\"a}llstr{\"o}min katu 2, University of Helsinki, Helsinki, Finland\label{aff89}
\and
Centre de Calcul de l'IN2P3/CNRS, 21 avenue Pierre de Coubertin 69627 Villeurbanne Cedex, France\label{aff90}
\and
Laboratoire d'etude de l'Univers et des phenomenes eXtremes, Observatoire de Paris, Universit\'e PSL, Sorbonne Universit\'e, CNRS, 92190 Meudon, France\label{aff91}
\and
SKA Observatory, Jodrell Bank, Lower Withington, Macclesfield, Cheshire SK11 9FT, UK\label{aff92}
\and
INFN-Sezione di Milano, Via Celoria 16, 20133 Milano, Italy\label{aff93}
\and
University of Applied Sciences and Arts of Northwestern Switzerland, School of Computer Science, 5210 Windisch, Switzerland\label{aff94}
\and
Universit\"at Bonn, Argelander-Institut f\"ur Astronomie, Auf dem H\"ugel 71, 53121 Bonn, Germany\label{aff95}
\and
Universit\'e C\^{o}te d'Azur, Observatoire de la C\^{o}te d'Azur, CNRS, Laboratoire Lagrange, Bd de l'Observatoire, CS 34229, 06304 Nice cedex 4, France\label{aff96}
\and
Universit\'e Paris Cit\'e, CNRS, Astroparticule et Cosmologie, 75013 Paris, France\label{aff97}
\and
CNRS-UCB International Research Laboratory, Centre Pierre Binetruy, IRL2007, CPB-IN2P3, Berkeley, USA\label{aff98}
\and
Institut d'Astrophysique de Paris, 98bis Boulevard Arago, 75014, Paris, France\label{aff99}
\and
Aurora Technology for European Space Agency (ESA), Camino bajo del Castillo, s/n, Urbanizacion Villafranca del Castillo, Villanueva de la Ca\~nada, 28692 Madrid, Spain\label{aff100}
\and
Institut de F\'{i}sica d'Altes Energies (IFAE), The Barcelona Institute of Science and Technology, Campus UAB, 08193 Bellaterra (Barcelona), Spain\label{aff101}
\and
DARK, Niels Bohr Institute, University of Copenhagen, Jagtvej 155, 2200 Copenhagen, Denmark\label{aff102}
\and
Waterloo Centre for Astrophysics, University of Waterloo, Waterloo, Ontario N2L 3G1, Canada\label{aff103}
\and
Department of Physics and Astronomy, University of Waterloo, Waterloo, Ontario N2L 3G1, Canada\label{aff104}
\and
Perimeter Institute for Theoretical Physics, Waterloo, Ontario N2L 2Y5, Canada\label{aff105}
\and
Centre National d'Etudes Spatiales -- Centre spatial de Toulouse, 18 avenue Edouard Belin, 31401 Toulouse Cedex 9, France\label{aff106}
\and
Institute of Space Science, Str. Atomistilor, nr. 409 M\u{a}gurele, Ilfov, 077125, Romania\label{aff107}
\and
Consejo Superior de Investigaciones Cientificas, Calle Serrano 117, 28006 Madrid, Spain\label{aff108}
\and
Universidad de La Laguna, Departamento de Astrof\'{\i}sica, 38206 La Laguna, Tenerife, Spain\label{aff109}
\and
Dipartimento di Fisica e Astronomia "G. Galilei", Universit\`a di Padova, Via Marzolo 8, 35131 Padova, Italy\label{aff110}
\and
INFN-Padova, Via Marzolo 8, 35131 Padova, Italy\label{aff111}
\and
Institut f\"ur Theoretische Physik, University of Heidelberg, Philosophenweg 16, 69120 Heidelberg, Germany\label{aff112}
\and
Institut de Recherche en Astrophysique et Plan\'etologie (IRAP), Universit\'e de Toulouse, CNRS, UPS, CNES, 14 Av. Edouard Belin, 31400 Toulouse, France\label{aff113}
\and
Universit\'e St Joseph; Faculty of Sciences, Beirut, Lebanon\label{aff114}
\and
Departamento de F\'isica, FCFM, Universidad de Chile, Blanco Encalada 2008, Santiago, Chile\label{aff115}
\and
Universit\"at Innsbruck, Institut f\"ur Astro- und Teilchenphysik, Technikerstr. 25/8, 6020 Innsbruck, Austria\label{aff116}
\and
Institut d'Estudis Espacials de Catalunya (IEEC),  Edifici RDIT, Campus UPC, 08860 Castelldefels, Barcelona, Spain\label{aff117}
\and
Satlantis, University Science Park, Sede Bld 48940, Leioa-Bilbao, Spain\label{aff118}
\and
Institute of Space Sciences (ICE, CSIC), Campus UAB, Carrer de Can Magrans, s/n, 08193 Barcelona, Spain\label{aff119}
\and
Department of Physics, Royal Holloway, University of London, TW20 0EX, UK\label{aff120}
\and
Instituto de Astrof\'isica e Ci\^encias do Espa\c{c}o, Faculdade de Ci\^encias, Universidade de Lisboa, Tapada da Ajuda, 1349-018 Lisboa, Portugal\label{aff121}
\and
Cosmic Dawn Center (DAWN)\label{aff122}
\and
Niels Bohr Institute, University of Copenhagen, Jagtvej 128, 2200 Copenhagen, Denmark\label{aff123}
\and
Universidad Polit\'ecnica de Cartagena, Departamento de Electr\'onica y Tecnolog\'ia de Computadoras,  Plaza del Hospital 1, 30202 Cartagena, Spain\label{aff124}
\and
Kapteyn Astronomical Institute, University of Groningen, PO Box 800, 9700 AV Groningen, The Netherlands\label{aff125}
\and
Infrared Processing and Analysis Center, California Institute of Technology, Pasadena, CA 91125, USA\label{aff126}
\and
Dipartimento di Fisica e Scienze della Terra, Universit\`a degli Studi di Ferrara, Via Giuseppe Saragat 1, 44122 Ferrara, Italy\label{aff127}
\and
Istituto Nazionale di Fisica Nucleare, Sezione di Ferrara, Via Giuseppe Saragat 1, 44122 Ferrara, Italy\label{aff128}
\and
INAF, Istituto di Radioastronomia, Via Piero Gobetti 101, 40129 Bologna, Italy\label{aff129}
\and
Instituto de Astrof\'isica de Canarias (IAC); Departamento de Astrof\'isica, Universidad de La Laguna (ULL), 38200, La Laguna, Tenerife, Spain\label{aff130}
\and
Universit\'e PSL, Observatoire de Paris, Sorbonne Universit\'e, CNRS, LERMA, 75014, Paris, France\label{aff131}
\and
Universit\'e Paris-Cit\'e, 5 Rue Thomas Mann, 75013, Paris, France\label{aff132}
\and
INAF - Osservatorio Astronomico di Brera, via Emilio Bianchi 46, 23807 Merate, Italy\label{aff133}
\and
INAF-Osservatorio Astronomico di Brera, Via Brera 28, 20122 Milano, Italy, and INFN-Sezione di Genova, Via Dodecaneso 33, 16146, Genova, Italy\label{aff134}
\and
ICL, Junia, Universit\'e Catholique de Lille, LITL, 59000 Lille, France\label{aff135}
\and
ICSC - Centro Nazionale di Ricerca in High Performance Computing, Big Data e Quantum Computing, Via Magnanelli 2, Bologna, Italy\label{aff136}
\and
Instituto de F\'isica Te\'orica UAM-CSIC, Campus de Cantoblanco, 28049 Madrid, Spain\label{aff137}
\and
CERCA/ISO, Department of Physics, Case Western Reserve University, 10900 Euclid Avenue, Cleveland, OH 44106, USA\label{aff138}
\and
Laboratoire Univers et Th\'eorie, Observatoire de Paris, Universit\'e PSL, Universit\'e Paris Cit\'e, CNRS, 92190 Meudon, France\label{aff139}
\and
Departamento de F{\'\i}sica Fundamental. Universidad de Salamanca. Plaza de la Merced s/n. 37008 Salamanca, Spain\label{aff140}
\and
Universit\'e de Strasbourg, CNRS, Observatoire astronomique de Strasbourg, UMR 7550, 67000 Strasbourg, France\label{aff141}
\and
Center for Data-Driven Discovery, Kavli IPMU (WPI), UTIAS, The University of Tokyo, Kashiwa, Chiba 277-8583, Japan\label{aff142}
\and
California Institute of Technology, 1200 E California Blvd, Pasadena, CA 91125, USA\label{aff143}
\and
Department of Physics \& Astronomy, University of California Irvine, Irvine CA 92697, USA\label{aff144}
\and
Department of Mathematics and Physics E. De Giorgi, University of Salento, Via per Arnesano, CP-I93, 73100, Lecce, Italy\label{aff145}
\and
INFN, Sezione di Lecce, Via per Arnesano, CP-193, 73100, Lecce, Italy\label{aff146}
\and
INAF-Sezione di Lecce, c/o Dipartimento Matematica e Fisica, Via per Arnesano, 73100, Lecce, Italy\label{aff147}
\and
Instituto de F\'isica de Cantabria, Edificio Juan Jord\'a, Avenida de los Castros, 39005 Santander, Spain\label{aff148}
\and
CEA Saclay, DFR/IRFU, Service d'Astrophysique, Bat. 709, 91191 Gif-sur-Yvette, France\label{aff149}
\and
Department of Computer Science, Aalto University, PO Box 15400, Espoo, FI-00 076, Finland\label{aff150}
\and
Instituto de Astrof\'\i sica de Canarias, c/ Via Lactea s/n, La Laguna 38200, Spain. Departamento de Astrof\'\i sica de la Universidad de La Laguna, Avda. Francisco Sanchez, La Laguna, 38200, Spain\label{aff151}
\and
Ruhr University Bochum, Faculty of Physics and Astronomy, Astronomical Institute (AIRUB), German Centre for Cosmological Lensing (GCCL), 44780 Bochum, Germany\label{aff152}
\and
Astrophysics Research Centre, University of KwaZulu-Natal, Westville Campus, Durban 4041, South Africa\label{aff153}
\and
School of Mathematics, Statistics \& Computer Science, University of KwaZulu-Natal, Westville Campus, Durban 4041, South Africa\label{aff154}
\and
Department of Physics and Astronomy, Vesilinnantie 5, 20014 University of Turku, Finland\label{aff155}
\and
Serco for European Space Agency (ESA), Camino bajo del Castillo, s/n, Urbanizacion Villafranca del Castillo, Villanueva de la Ca\~nada, 28692 Madrid, Spain\label{aff156}
\and
ARC Centre of Excellence for Dark Matter Particle Physics, Melbourne, Australia\label{aff157}
\and
Centre for Astrophysics \& Supercomputing, Swinburne University of Technology,  Hawthorn, Victoria 3122, Australia\label{aff158}
\and
Department of Physics and Astronomy, University of the Western Cape, Bellville, Cape Town, 7535, South Africa\label{aff159}
\and
DAMTP, Centre for Mathematical Sciences, Wilberforce Road, Cambridge CB3 0WA, UK\label{aff160}
\and
Kavli Institute for Cosmology Cambridge, Madingley Road, Cambridge, CB3 0HA, UK\label{aff161}
\and
Department of Astrophysics, University of Zurich, Winterthurerstrasse 190, 8057 Zurich, Switzerland\label{aff162}
\and
IRFU, CEA, Universit\'e Paris-Saclay 91191 Gif-sur-Yvette Cedex, France\label{aff163}
\and
Oskar Klein Centre for Cosmoparticle Physics, Department of Physics, Stockholm University, Stockholm, SE-106 91, Sweden\label{aff164}
\and
Astrophysics Group, Blackett Laboratory, Imperial College London, London SW7 2AZ, UK\label{aff165}
\and
Univ. Grenoble Alpes, CNRS, Grenoble INP, LPSC-IN2P3, 53, Avenue des Martyrs, 38000, Grenoble, France\label{aff166}
\and
INAF-Osservatorio Astrofisico di Arcetri, Largo E. Fermi 5, 50125, Firenze, Italy\label{aff167}
\and
Dipartimento di Fisica, Sapienza Universit\`a di Roma, Piazzale Aldo Moro 2, 00185 Roma, Italy\label{aff168}
\and
Centro de Astrof\'{\i}sica da Universidade do Porto, Rua das Estrelas, 4150-762 Porto, Portugal\label{aff169}
\and
Dipartimento di Fisica, Universit\`a di Roma Tor Vergata, Via della Ricerca Scientifica 1, Roma, Italy\label{aff170}
\and
INFN, Sezione di Roma 2, Via della Ricerca Scientifica 1, Roma, Italy\label{aff171}
\and
HE Space for European Space Agency (ESA), Camino bajo del Castillo, s/n, Urbanizacion Villafranca del Castillo, Villanueva de la Ca\~nada, 28692 Madrid, Spain\label{aff172}
\and
Dipartimento di Fisica - Sezione di Astronomia, Universit\`a di Trieste, Via Tiepolo 11, 34131 Trieste, Italy\label{aff173}
\and
Department of Astrophysical Sciences, Peyton Hall, Princeton University, Princeton, NJ 08544, USA\label{aff174}
\and
Theoretical astrophysics, Department of Physics and Astronomy, Uppsala University, Box 515, 751 20 Uppsala, Sweden\label{aff175}
\and
Minnesota Institute for Astrophysics, University of Minnesota, 116 Church St SE, Minneapolis, MN 55455, USA\label{aff176}
\and
Mathematical Institute, University of Leiden, Einsteinweg 55, 2333 CA Leiden, The Netherlands\label{aff177}
\and
Institute of Astronomy, University of Cambridge, Madingley Road, Cambridge CB3 0HA, UK\label{aff178}
\and
Space physics and astronomy research unit, University of Oulu, Pentti Kaiteran katu 1, FI-90014 Oulu, Finland\label{aff179}
\and
Department of Physics and Astronomy, Lehman College of the CUNY, Bronx, NY 10468, USA\label{aff180}
\and
American Museum of Natural History, Department of Astrophysics, New York, NY 10024, USA\label{aff181}
\and
Center for Computational Astrophysics, Flatiron Institute, 162 5th Avenue, 10010, New York, NY, USA\label{aff182}
\and
Department of Physics and Astronomy, University of British Columbia, Vancouver, BC V6T 1Z1, Canada\label{aff183}}    

%
%
%
%

%
%

%
\abstract{
Strong gravitational lensing systems with multiple source planes are powerful tools for probing the density profiles and dark matter substructure of the galaxies. The ratio of Einstein radii is related to the dark energy equation of state through the cosmological scaling factor $\beta$. However, galaxy-scale double-source-plane lenses (DSPLs) are extremely rare. In this paper, we report the discovery of four new galaxy-scale double-source-plane lens candidates in the Euclid Quick Release 1 (Q1) data. These systems were initially identified through a combination of machine learning lens-finding models and subsequent visual inspection from citizens and experts. We apply the widely-used {\tt LensPop} lens forecasting model to predict that the full \Euclid survey will discover 1700 DSPLs, which scales to $6 \pm 3$ DSPLs in 63 deg$^2$, the area of Q1. The number of discoveries in this work is broadly consistent with this forecast. We present lens models for each DSPL and infer their $\beta$ values. Our initial Q1 sample demonstrates the promise of \Euclid to discover such rare objects.}
%
%
%
   \titlerunning{Double-source-plane Lenses in Euclid Q1}
    \authorrunning{Euclid Collaboration: Li et al}
   
 \maketitle
%
%
%
%
\section{Introduction}

Strong gravitational lensing occurs when a background source and a foreground massive object align along our line of sight, resulting in the light from the source being split into multiple images or forming an Einstein ring \citep{Einstein1936, Zwicky1937a, Zwicky1937b}. Sometimes there can be multiple sources at different redshifts behind the same lens galaxy, forming Einstein rings at different radii. Such systems are referred to as compound lenses or double-source-plane lenses (DSPLs). 

In strong lensing systems, the radius of the Einstein ring, known as the Einstein radius (parameterised by $\theta_\mathrm{E}$), depends on both the mass profile of the lens and the cosmological distances involved. Performing lens modelling on systems with a single Einstein ring can constrain the local logarithmic density slope in the region where the strongly lensed images are observed, assuming a specific parametric form for the total mass profile of the lens galaxy \citep[e.g.,][]{Suyu2010, Birrer2018, Nightingale2021, Galan2022}. However, there are systematics affecting the results, such as the mass-sheet degeneracy \citep[]{Falco1985, Schneider:2013sxa}: by adding a constant mass sheet and rescale the convergence, mass-sheet transformations (MST) can alter the shape of the galaxy mass profile and rescaling the size of the source while keeping all lensing observables unchanged. In DSPL systems, adding an extra source plane can break the mass-sheet degeneracy of the lens if the mass contribution of the first source is neglected \citep{Brada2004} and the cosmology is known. 

The presence of two rings at different radii makes DSPLs powerful tools for probing the mass distribution of galaxies, thereby constraining the properties of dark matter. The second ring provides an additional aperture within which the total mass is well-constrained, so one can disentangle the distribution of dark and luminous matter in the lens without additional kinematic data \citep{Sonnenfeld_2012}. If a dark matter subhalo exists in the lens plane, the inclusion of a second source probes the mass distribution on a larger radius. This can break degeneracies in the lens model and improve the constraints on its mass properties \citep[see Figure 6 in][ where the posteriors of the dark matter subhalo when modelled with two arcs are tighter]{Enzi2024}. For example, several observations and analyses of the `Jackpot' lens SDSS J0946+1006 have suggested the existence of an over-concentrated subhalo \citep[see][]{vegetti2010detection, Minor:2020hic, Ballard:2023fgi, Despali:2024ihn, Enzi2024}. 

DSPLs have also been used to constrain the equation of state of dark energy. For a system with multiple sources at different redshifts, the ratio of their Einstein radii is related to cosmological parameters, such as the dark energy equation of state (parameterised by $w$), through the cosmology scaling factor $\beta$. The cosmological measurements derived from DSPLs are independent of the Hubble constant and complement other cosmological measurements, as demonstrated by previous studies \citep[see e.g.,][]{Gavazzi2008, collett2012, Johnson2025}. \cite{Collett2014} shows that a single galaxy-scale DSPL system can provide competitive cosmological constraints assuming a power-law mass profile of the lens galaxy. Although cluster-scale lenses typically have multiple source planes and are also capable of measuring cosmological parameters \citep[e.g.,][]{Soucail2004, Jullo2010, Caminha2016, Acebron:2017mim, Magana:2017gfs,Caminha2022}, their complex mass distributions limit the precision. While selecting strong lensing clusters or galaxy groups with simpler mass distributions is a possible alternative \citep{Bolamperti2024}, galaxy-scale lenses generally exhibit less systematic uncertainty due to their inherently simpler mass profiles. However, \citet{Schneider2014} argues that there exists an analogue of the MST for DSPL systems, which can render the cosmological constraints significantly less restrictive. A 1 percent bias in $\beta$ can give rise to a bias in $w$ of approximately 0.3. Additional data from stellar kinematics, for example, can help break this degeneracy; however, achieving competitive results requires unbiased measurements of kinematics and correct assumptions about the mass profiles -- both of which are challenging.

\begin{figure*}
    \centering
    \includegraphics[width=0.245\textwidth]{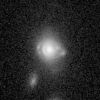}
    \includegraphics[width=0.245\textwidth]{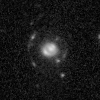}
    \includegraphics[width=0.245\textwidth]{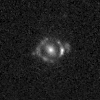}
    \includegraphics[width=0.245\textwidth]{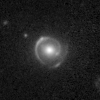}

    \includegraphics[width=0.245\textwidth]{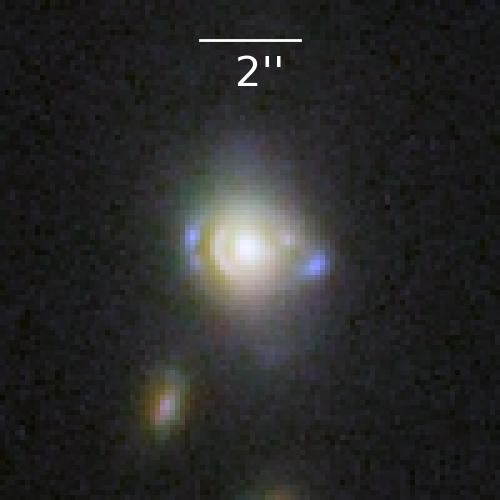}
    \includegraphics[width=0.245\textwidth]{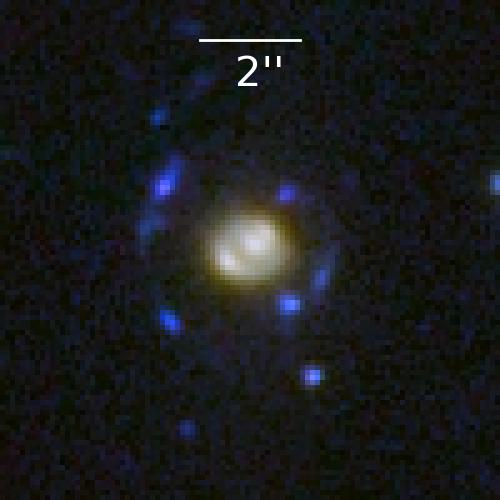}
    \includegraphics[width=0.245\textwidth]{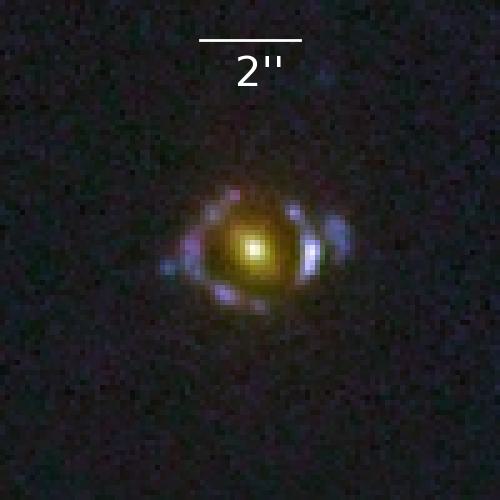}
    \includegraphics[width=0.245\textwidth]{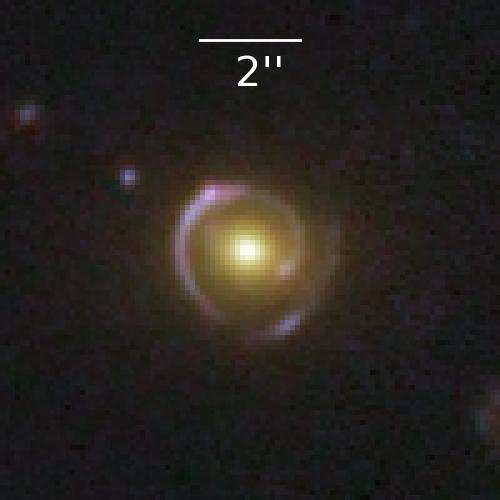}
    
    \caption{\IE (top panels) and \IE-\YE-\JE coloured images (bottom panels) of the DSPL candidates. From left to right there are the \tpl, \cdl,  \gsl, and \ssl, respectively. Orientation: north is up, and east is to the left.}
    \label{fig:images}
\end{figure*}

Unfortunately, galaxy-scale strong lenses with multiple background sources are exceedingly rare, with only a handful of known examples in the literature: the `Jackpot' lens \citep{Gavazzi2008}; the `Eye of Horus' \citep{Tanaka2016}; and J1721+8842 \citep{Dux2024}. Other DSPLs which the sources have similar Einstein radii are: DES0408$-$5354 \citep{2017ApJ...838L..15L}; and the ‘Cosmic horseshoe’ \citep{2007ApJ...671L...9B}. This rarity arises from the fact that whilst the probability of being a single plane lens scales as $\theta_{\rm E}^2$, the probability of lensing two sources scales as $\theta_{\rm E}^4$. Even with the depth and resolution of the \emph{Hubble} Space Telescope, only about one in a hundred lenses are expected to be DSPL \citep{Gavazzi2008}, although the exact rate depends on the depth, resolution, and wavelength of the observations \citep{collett2012}. With the \Euclid space telescope providing an unprecedented inspection of the sky, forecasts shows that around 1700 DSPLs will be discovered during the survey \citep[rescaling from the \num{170000} single-source-plane lenses forecast in][]{Collett:2015roa}.

\begin{table}[]
    \centering
    \renewcommand{\arraystretch}{1.5}
    \caption{Sky locations are RA and Dec of the four lens systems}
    \begin{tabular}{lrr}
    \hline
    \hline
        Name & RA & Dec\\
        \hline
        \tpl & 273.595\degree &  67.138\degree\\
        \cdl & 59.565\degree &  $-$50.948\degree\\
        \gsl&  66.685\degree & $-$48.111\degree\\
        \ssl&  61.239\degree & $-$49.373\degree\\
        \hline
    \end{tabular}
    \label{tab:lens}
\end{table}

Here we report the discovery of four new galaxy-scale DSPL candidates in the 63 deg$^2$ of Euclid Quick Release 1 \citep[Q1][]{Q1cite}. The \IE-band images and coloured images of the four systems are presented in Fig. \ref{fig:images}, and their coordintes are shown in Table \ref{tab:lens}. Based on the morphology of the four discovered lens systems, they have been named the `\tpl', `\cdl', `\gsl'$\text{ }$(this lensing configuration somewhat resembles Galileo's hand-drawn depiction of Saturn), and `\ssl'.

The discovery of these four new DSPLs were made using the Strong Lensing Discovery Engine, a structured workflow that integrates machine learning (ML) with citizen science classifications, followed by expert grading. This paper is part of a series of papers, with \citet{Q1-SP048} describing the \Euclid lens finding engine and new discoveries. \citet{Q1-SP052} describes our effort to build a suitable training sample for Q1 and the lens search with high-velocity dispersion galaxies. \citet{Q1-SP053} explains the ML models used to find the strong lenses. Finally \cite{Q1-SP059} introduces an ensemble classifier combining the citizen science and the different ML classifiers. 

The paper is organised as follows. In Sect. \ref{sec:Euclid}, we briefly summarise the \Euclid survey and the lens-finding methods employed. Section \ref{sec:modelling} introduces the lens modelling algorithm, followed by the lens models and spectra presented in Sect. \ref{sec:result}. In Sect. \ref{sec:catalog} we present the forecast of the number of DSPLs that \Euclid will discover in its planned survey area. Finally, Sect. \ref{sec:conclusion} presents our conclusions. We discuss the limitations of our lens-finding approach and provide examples of typical false DSPL systems in Appendix. \ref{sec:imposter}.

\begin{figure*}
    \centering
    \includegraphics[width=0.95\textwidth]{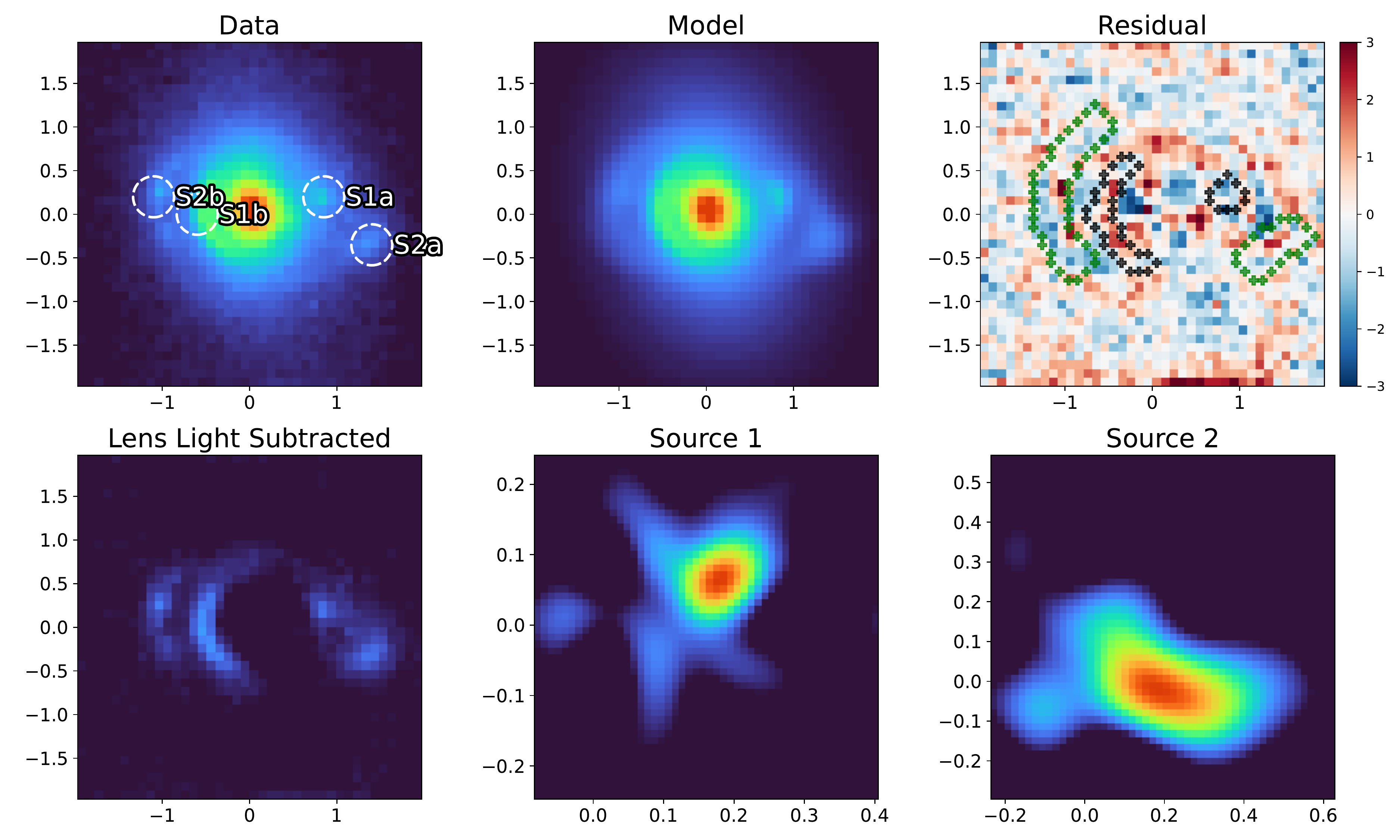}
    \caption{Lens model of the \tpl. \IE band image, lens model, and the residual respectively. The bottom panels display the lens light-subtracted image and two source models. The inner pair of arcs are labelled as S1a and S1b, while the outer pair of arcs are labelled as S2a and S2b. In the residual plot, the black and green scatter points represent the edges of the masks for the inner and outer arcs, respectively. The axis tick marks are in arcseconds.}
    \label{fig:Natlens}
\end{figure*}

\section{\Euclid double-source-plane lens search}
\label{sec:Euclid}

The search for DSPLs is a by-product of the Strong Lensing Discovery Engine, which had a primary goal of finding single-source-plane galaxy-scale strong lenses in \Euclid data. In the lens-finding pipeline described in \citet{Q1-SP048}, the lens candidates selected by machine learning and citizen scientists (from objects that are brighter than 22.5 in \IE in the 63 deg$^2$ of Q1) are sent to the Galaxy Judges project, where a group of strong lensing experts provide a final grade for each lens candidate. Following this semi-automated search for all galaxy-galaxy lenses, a group of (approximatly 10) experienced experts re-inspected the top \num{10000} lenses, and identified four DSPLs by eye. All four of our DSPL candidates are ranked in the top 250 lenses by the expert visual inspection.

The primary criterion for identifying a DSPL system is the presence of two pairs of arcs at different Einstein radii. The colours of the arcs serve as a key indicator to determine whether they originate from the same source plane. Once a lens is identified as a potential DSPL candidate, preliminary lens modelling (described in Sect. \ref{sec:modelling}) is performed to further evaluate the plausibility of its configuration. 

\Euclid offers significant advantages for DSPL searches due to its high resolution imaging over a large area, with its point-spread function (PSF) having a full width at half maximum of approximately $\ang{;;0.16}$ in the \IE band \citep{EuclidSkyOverview, Q1-TP002}. This high resolution enables the separation of lens light from the arcs and allows multiple arcs to be resolved. Although the infrared channels have a much lower resolution, their colour information is helpful for associating different arcs.

\begin{figure*}
    \centering
    \includegraphics[width=0.95\textwidth]{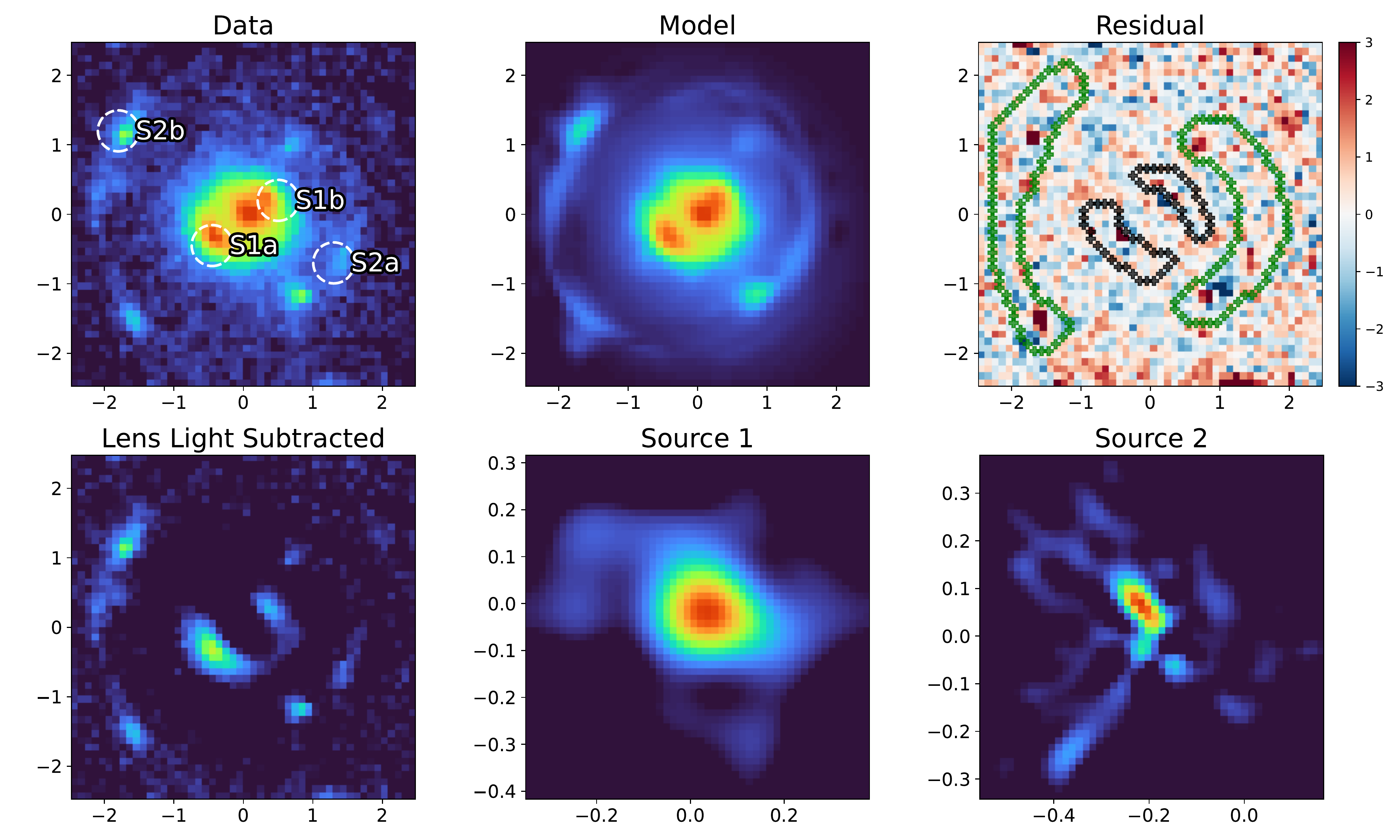}
    \caption{Same as Fig. \ref{fig:Natlens}, but for the \cdl.}
    \label{fig:tomlens}
\end{figure*}

\section{Double-source-plane lens modelling}
\label{sec:modelling}

\subsection{Background theory}

For a lensed image at position $\boldsymbol{\theta}$, the scaled deflection angle of a galaxy $\boldsymbol{\alpha}(\boldsymbol{\theta})$ is related to its lensing potential $\psi$ via
\begin{equation}
\boldsymbol{\alpha}(\boldsymbol{\theta})=\nabla \psi(\boldsymbol{\theta})\,,
\end{equation}
and the relation between lensing potential and lensing convergence is
\begin{equation}
\kappa(\boldsymbol{\theta})=\frac{1}{2} \nabla^2 \psi(\boldsymbol{\theta})\,,
\end{equation}
where convergence is defined as,
\begin{equation}
\kappa(\boldsymbol{\theta}) \equiv \frac{\Sigma(\boldsymbol{\theta})}{\Sigma_{\mathrm{cr}}}\,.
\end{equation}
The convergence is the surface mass density normalised by the critical lensing surface density
\begin{equation}
\Sigma_{\mathrm{cr}} \equiv \frac{c^2 D_{\mathrm{s}}}{4 \pi G D_{\mathrm{l}} D_{\mathrm{ls}}}\,,
\end{equation}
where \(D\) is the angular diameter distance between two objects, and the subscripts `l' and `s' denote the lens galaxy and the source galaxy, respectively.

In a DSPL system, the lens equation of the first source plane can be written as
\begin{equation}
\mathbf{\boldsymbol{\theta}}_1 = \mathbf{\boldsymbol{\theta}}-\beta\, \boldsymbol{\alpha}_1(\mathbf{\boldsymbol{\theta}})\,,
\end{equation}
where \({\boldsymbol{\alpha}_1}\) is the reduced deflection angle of the first source at the image position $\boldsymbol{\theta}$, and \(\boldsymbol{\theta}_1\) is the position of the first source on its source plane. Here, \(\beta\) is the cosmological scaling factor which is the ratio of the angular diameter distances between the different redshift planes:
\begin{equation}
\label{eq:beta}
\beta = \frac{D_\mathrm{ls1} D_\mathrm{s2}}{D_\mathrm{s1} D_\mathrm{ls2}}\,,
\end{equation}
where the subscripts `s1' and `s2' denote source 1 and source 2, respectively. The lens equation of the second source plane is then:
\begin{equation}
\boldsymbol{\theta}_2 = \boldsymbol{\theta}-\boldsymbol{\alpha}_1(\boldsymbol{\theta})-\boldsymbol{\alpha}_\mathrm{s1}[\boldsymbol{\theta}-\beta\,\boldsymbol{\alpha}_1(\boldsymbol{\theta})]\,,
\end{equation}
where $\boldsymbol{\theta}_2$ is the position of the second source on its source plane and $\boldsymbol{\alpha}_\mathrm{s1}$ is the deflection angle of the mass on the first source plane. In a simple case where we assume the lens galaxy has a singular isothermal sphere (SIS) profile, the parameter \(\beta\) is just the ratio of the two Einstein radii \citep[the geometry of the DSPL is in Figure 1 of][]{Collett2014}. 

The lenses are modelled using an elliptical power-law (EPL) mass model with an external shear component. The convergence ($\kappa_{\mathrm{EPL}}$) of the EPL model can be parameterised as
\begin{equation}
\kappa_{\mathrm{EPL}}(\rho, \gamma, q)=\frac{3-\gamma}{2}\left(\frac{\theta_\mathrm{E}}{\rho}\right)^{\gamma-1}\,,
\end{equation}
where $\rho^2=x^2 q+y^2 / q$ is the squared elliptical radius and $q$ is the axis ratio, $\gamma$ is the power-law index of the density slope, and $\theta_\mathrm{E}$ is the Einstein radius corresponding to source 2. The cosmological parameters can therefore be measured by relating $\beta$ from lens modelling and the redshifts of the system.

\subsection{Lens modelling strategy}

We use the open-source lens modelling code {\tt Herculens}\footnote{\url{https://github.com/Herculens/herculens}} \citep{Galan2022_herculens} in combination with a multiplane extension. {\tt Herculens} is based on the automatic differentiation and compilation features of JAX\footnote{\url{https://docs.jax.dev/en/latest/}} and can run on graphics processing units (GPUs). The detailed description of the extension can be found in \cite{Enzi2024}. We note here that our lens modelling is primarily aiming to the validate the plausibility of these systems being DSPLs, and not to precisely infer the parameters of these systems. The lack of redshift information for either the lens or the sources prevents us from building robust mass distribution and leverage the full statistical analyses from Markov chain Monte Carlo (MCMC) methods. However, our lens modelling is fully capable of recovering the source-plane morphology of the galaxies in order to confirm the double-source-plane nature of the systems. Here we briefly summarise the modelling strategy. 

The data used for modelling comes from the \IE band because it has the highest pixel resolution ($0\farcs1$ pixel$^{-1}$). We measure the root mean square of the background pixels and infer the shot noise from the modelled lens light. The PSF is derived from the \Euclid pipeline. Since the \Euclid \IE band spans a wide wavelength range (550--900 nm), the PSF profile will be influenced by the spectral energy distribution of the target. Consequently, lensed arcs with markedly different colours might exhibit different PSFs, which might introduce additional uncertainties into the lens model.

We use {\tt Numpyro}’s implementation of stochastic variational inference \citep[SVI; see][]{WingateWeber2013}. We employ the AdaBelief optimiser \citep{Zhuang2020} with a low-rank multivariate normal distribution as the guiding probability distribution. This guiding distribution is not complex enough to accurately recover the full posterior distribution, but it is computationally much cheaper than MCMC methods because it can use the auto-differentiable nature of JAX \citep{jax2018github}. The inference was done through optimising the guiding distribution to minimise the Kullback--Leibler divergence to the true posterior rather than sampling over the posterior.

\begin{figure*}
    \centering
    \includegraphics[width=0.95\textwidth]{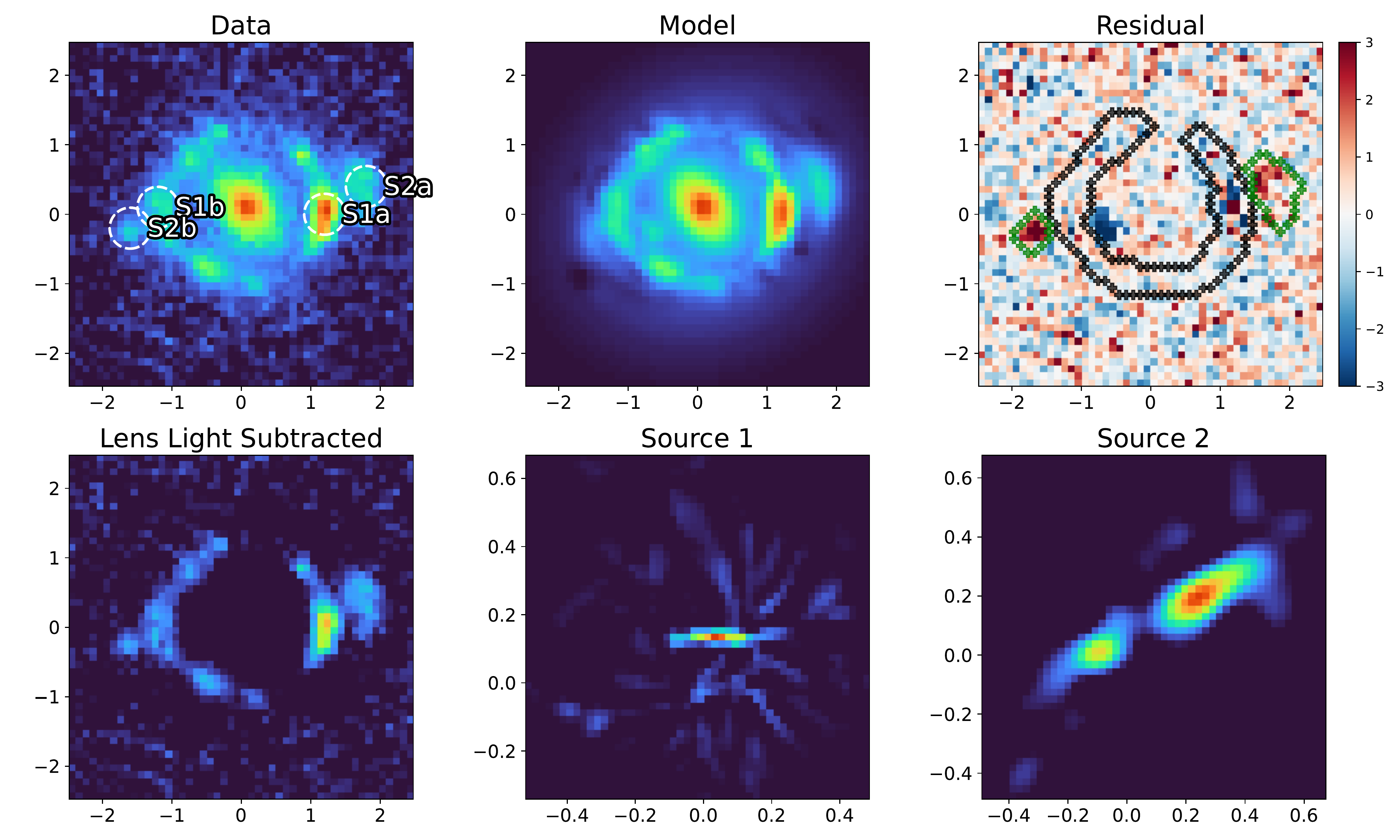}
    \includegraphics[width=0.95\textwidth]{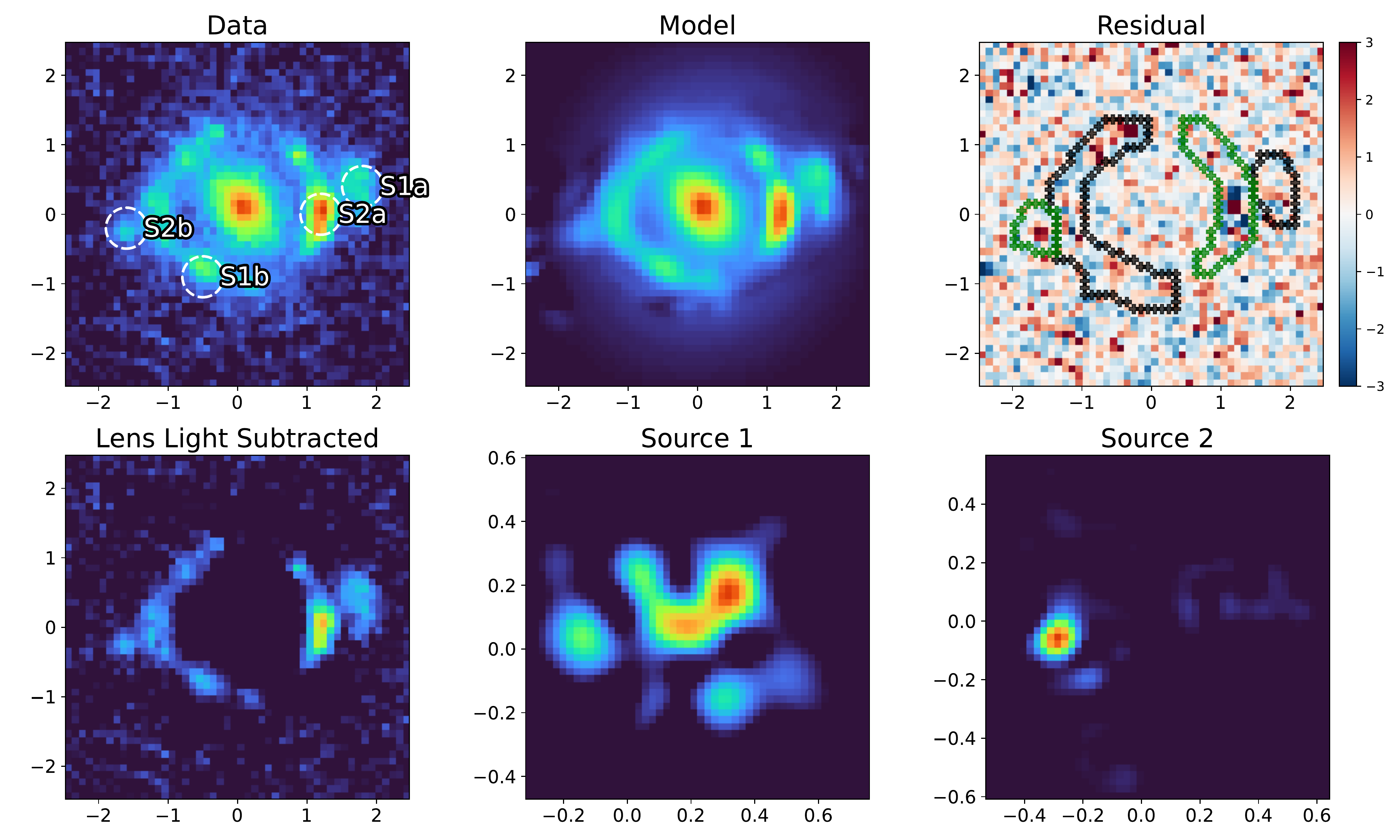}
    \caption{Same as Fig. \ref{fig:Natlens} but for \gsl. The top six figures are the lens modelled with a single source plane. The bottom six figures are the lens modelled with double-source-planes.}
    \label{fig:noneimposter}
\end{figure*}

\begin{figure*}
    \centering
    \includegraphics[width=0.95\textwidth]{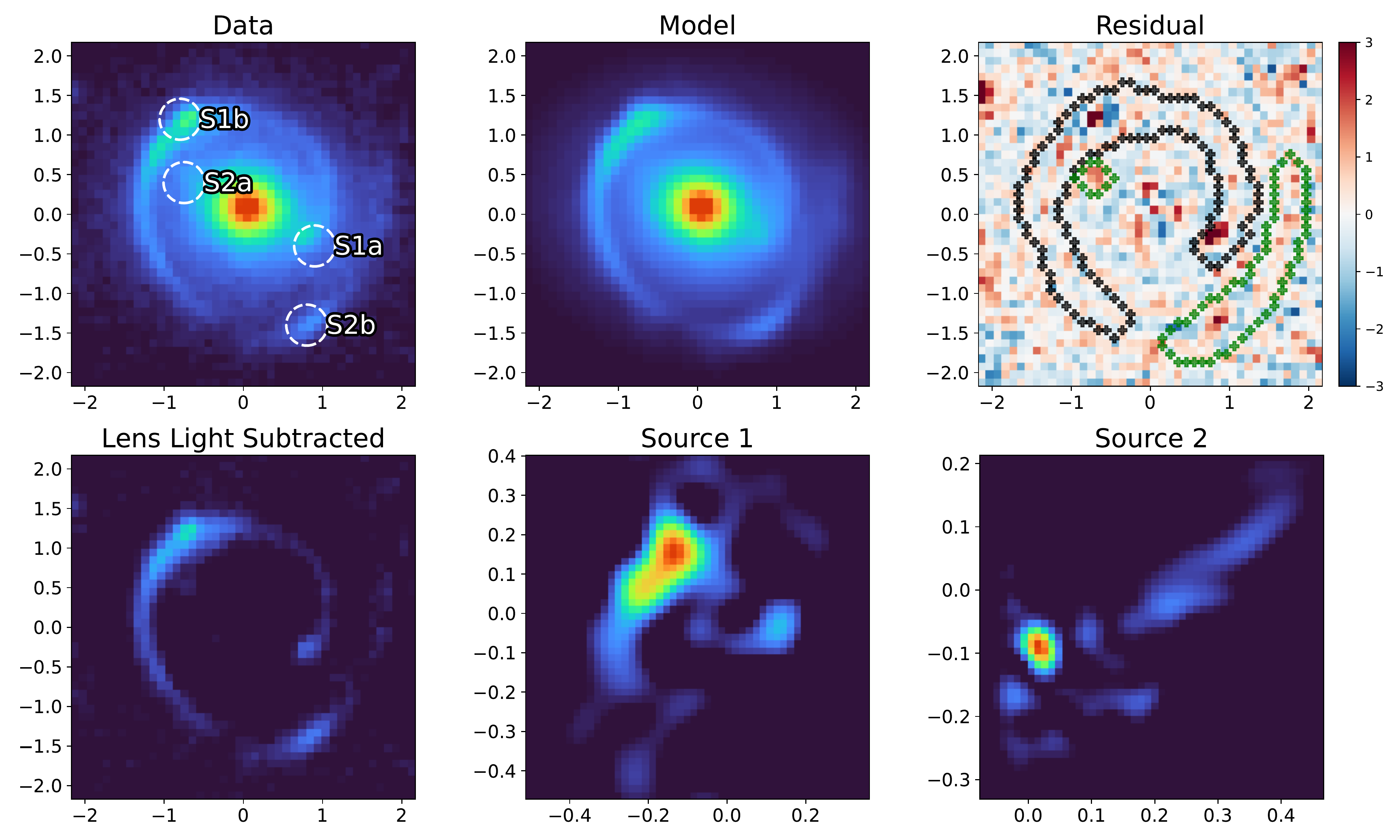}
    \caption{Same as Fig. \ref{fig:Natlens}, but for \ssl.}
    \label{fig:snail}
\end{figure*}

\begin{figure*}
    \centering
    \includegraphics[width=0.95\textwidth]{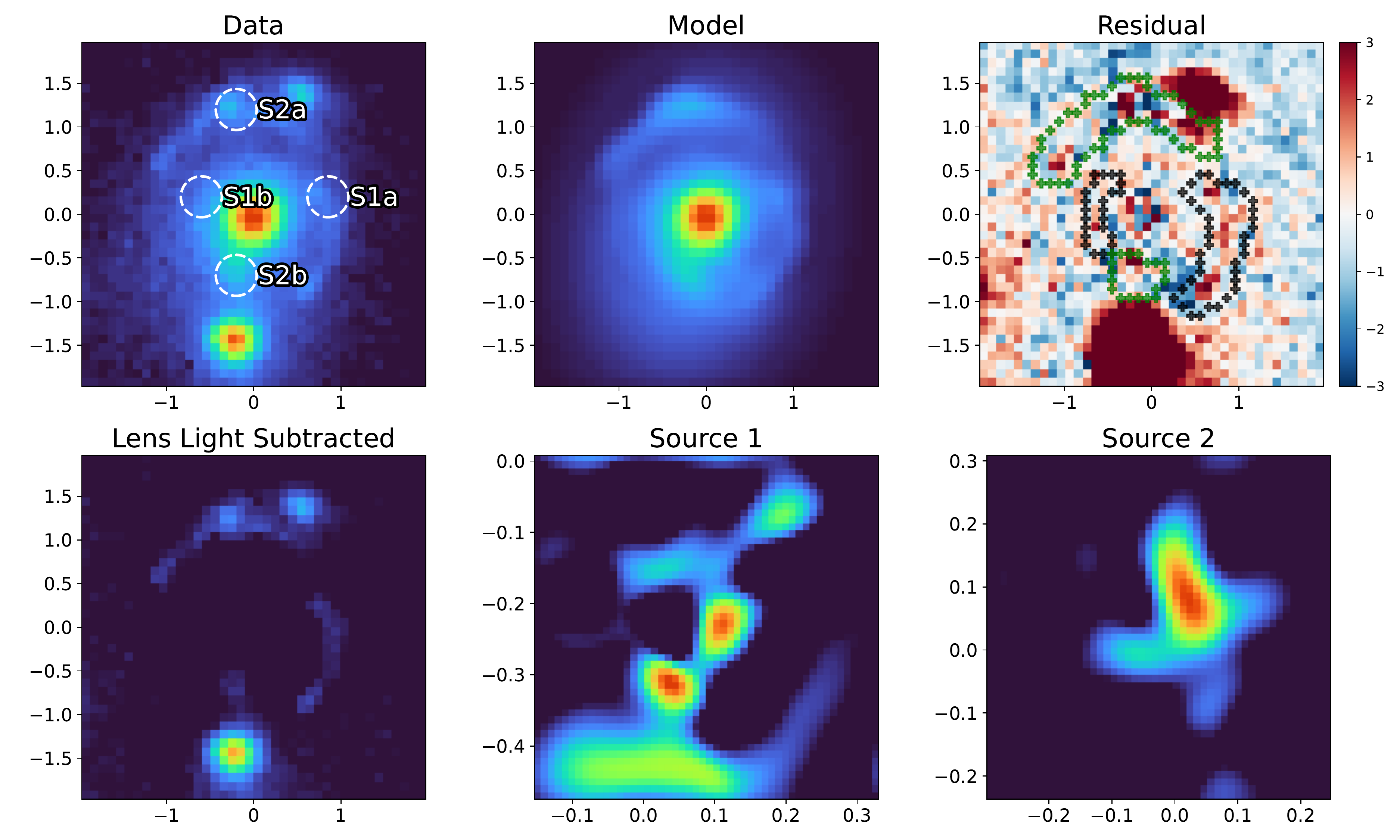}
    \caption{Same as Fig. \ref{fig:Natlens}. But for the CrackPot lens.}
    \label{fig:EDFN10}
\end{figure*}

Since the inner arcs of our DSPL candidates are mixed with the lens light, we model the lens light simultaneously with the arcs. In our model, the lens light is modelled with the sum of five elliptical Gaussian luminosity profiles with flexible centres \citep{Shajib2019, He2024, Enzi2024}. The sources are first modelled with a single elliptical Gaussian luminosity profile to provide an approximate answer. Then we use the parameterised lens model as the starting point and describe the sources as fields that vary on regularly pixelated grids. We compute each pixel’s field value by multiplying the Matérn power spectrum \citep[see][]{stein2012} with white noise (sampled from a standard normal distribution) and then applying the inverse Fourier transform to the result. A flat prior has been applied to every mass model parameter.

For each of our targets, we run \num{20000} iterations in 10 SVIs. Each SVI realisation has different random initialisations, and we pick the SVI result with the smallest average loss value to be our lens model. The lens model results are shown in Table \ref{tab:result}. We do not show the uncertainty of the lens model because SVI tends to underestimate the lens model uncertainty. Uncertainties from the following MCMCs are typically to be around 0.05 for each parameter between the 10 SVI runs.

\section{Lens model of DSPL candidates}

In this section, we present and analyse the lens models of our DSPL candidates (\tpl, \cdl, \gsl, and \ssl). We also present one system that we initially thought was a DSPL candidate, the 'Crackpot', which was found to be better modelled with two sources on a single plane (Sect \ref{sec:crackpot}). Since the Crackpot is not a DSPL lens, it is not shown in Table \ref{tab:result}.

\label{sec:result}
\subsection{The \tpl}
This lens was discovered during the visual inspection of the top \num{20000} ranked lenses according to the lens-finding version of the machine-learning model {\tt Zoobot} \citep[see][]{Q1-SP053}. It ranked 2349 by {\tt Zoobot} and 43 in Galaxy Judges. It was classified as a `Grade A' lens. The system features two concentric rings at distinct Einstein radii, each with a different colour, making it a highly promising DSPL candidate.

The lens model of \tpl$\text{ }$is presented in Fig. \ref{fig:Natlens}. The lens light-subtracted image reveals a distinct pair of inner rings. The lens model results indicate an inner ring at $0\farcs62$ and a $\beta$ of 0.74, leaving a clean residual within the arc's position. The positive residual near the centre could be due to multiple reasons: dust lanes or AGN in the lens galaxy; PSF mismatch impacting the lens light subtraction; a satellite galaxy located close to the primary lens; or an inner 'zig-zag' image of the second source \citep{collettbacon2016}.

\subsection{\cdl}
This lens was discovered during the visual inspection of the top \num{20000} ranked lenses. It ranked 1902 by {\tt Zoobot} and 94 in Galaxy Judges. It was classified as a 'Grade A' lens.
The lens model of Cosmic Dartboard is presented in Fig. \ref{fig:tomlens}. Our lens model successfully recovered all the details in the image. However, due to limited resolution, it is challenging to perfectly disentangle the lens light from the inner arc. The outer arc reconstruction could be significantly improved with higher signal-to-noise ratio data.

The inner ring of this lens was inferred to be at $0\farcs45$, while the $\beta$ value is 0.51. The first source is massive according to the result from the lens model ($\theta_\mathrm{sis1} = 0\farcs48$) because it appears to be an elliptical galaxy. If spectroscopic redshifts are obtained, the large Einstein radius ratio should provide valuable insights for cosmological measurements as the image configuration suggests $z_\mathrm{s1} \ll z_\mathrm{s2}$, which would provide the optical lever arm for constraining dark energy \citep{collett2012}.  

\subsection{\gsl}

This lens was first discovered in Space Warps and Galaxy Judges, it ranked 3 by {\tt Zoobot} (though it was missed as a DSPL candidate in the first pass of {\tt Zoobot} by TC, TL and NL) and 19 in Galaxy Judges. It was classified as a 'Grade A' lens.
This system was proposed as a DSPL system because it features a central ring structure and two faint arcs on the outskirts. However, the apprent inner ring has an unusual configuration and the colours of all the arcs are similar, making it difficult to classify confidently. It is also difficult to explain both the very elliptical central lens light and the nearly circular ring merely by classifying it as a ring galaxy.

Here, we present two lens models. The first model assumes a DSPL configuration, while the second model considers a lens with two sources situated on the same redshift plane. Neither model can explain the lensing configuration really well. It is possible that this system is a single ring system where the two blue objects nearby are just field galaxies. Spectroscopic data or imaging with higher resolution should provide definitive evidence to decide whether this system is indeed a DSPL.

\subsection{\ssl}

This lens was discovered during the visual inspection of the top 1000 Galaxy Judges candidates. It ranked 7 in {\tt Zoobot} (but was initially assumed to be a multicomponent single source plane system) and 207 in Galaxy Judges. It was classified as a 'Grade B' lens, but the expert scores were not unanimous, with 4 A* votes (lens with special interest), 1 A, 3 B, and 2 X (non-lens).

The lens model of \ssl ~is presented in Fig.~\ref{fig:snail}. Our lens model successfully recovered all the details in the image. The $\beta$ of this system is constrained to be around 0.85. Although this system was modelled with excellent fidelity and is almost certainly a DSPL, it received a surprisingly low score in Galaxy Judges. The best explanation is that the lensing configuration of this system is uncommon, so human experts are less confident about this object. We initially modelled this system expecting to rule it out as a DSPL, with the two sources on a single plane, however the modelling result shows that this is a confident DSPL with $\beta \approx 0.85$. This demonstrates the power of lens modelling over human experts in identifying DSPLs with $\beta$ not much smaller than 1.

\subsection{The Crackpot: a single plane system with two sources}
\label{sec:crackpot}

\begin{figure}
    \centering
    \includegraphics[width=0.35\textwidth]{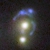}
    \caption{The colour image of the Crackpot lens. }
    \label{fig:crackpot}
\end{figure}



\begin{table*}
    \centering
    \renewcommand{\arraystretch}{1.5}
    \caption{Properties and lens model parameters of the four systems. We show the lensed \IE-band magnitudes of the lens, source 1 (s1), and source 2 (s2). The lens model parameters are the best solutions after 10 SVI runs per lens. Uncertainties are typically around 0.05 for each parameter between the 10 SVI runs.}
    \begin{tabular}{lcccccccccc}
    \hline
    \hline
        Name & $m_\mathrm{ab}$(lens) (\IE) & $m_\mathrm{ab}$(s1) (\IE) & $m_\mathrm{ab}$(s2) (\IE) & $\beta$ & $\theta_{\rm E1}$  & $\theta_{\rm sis1}$ & $\theta_{\rm E2}$ & $\gamma$ & $\phi$ & $q$ \\
        \hline
        \tpl & 20.0 & 23.3 & 24.0 & 0.74 & \ang{;;0.62} & \ang{;;0.13} & \ang{;;0.89} & 1.83 & $-$0.44 & 0.63 \\
        \cdl & 21.2 & 22.9 & 23.2 & 0.51 & \ang{;;0.45} & \ang{;;0.48} & \ang{;;1.31} & 1.63 & $-$1.32 & 0.64 \\
        \gsl & 22.3 & 22.7 & 24.4 & 0.69 & \ang{;;1.09} & \ang{;;0.28} & \ang{;;1.52} & 2.11 & \phantom{$-$}0.39 & 0.76 \\
        \ssl & 20.0 & 22.3 & 24.2 & 0.85 & \ang{;;1.18} & \ang{;;0.15} & \ang{;;1.35} & 2.22 & \phantom{$-$}0.17 & 0.72 \\
        \hline
    \end{tabular}
    \label{tab:result}
\end{table*}

\begin{figure*}
    \centering
    \includegraphics[width=0.9\textwidth]{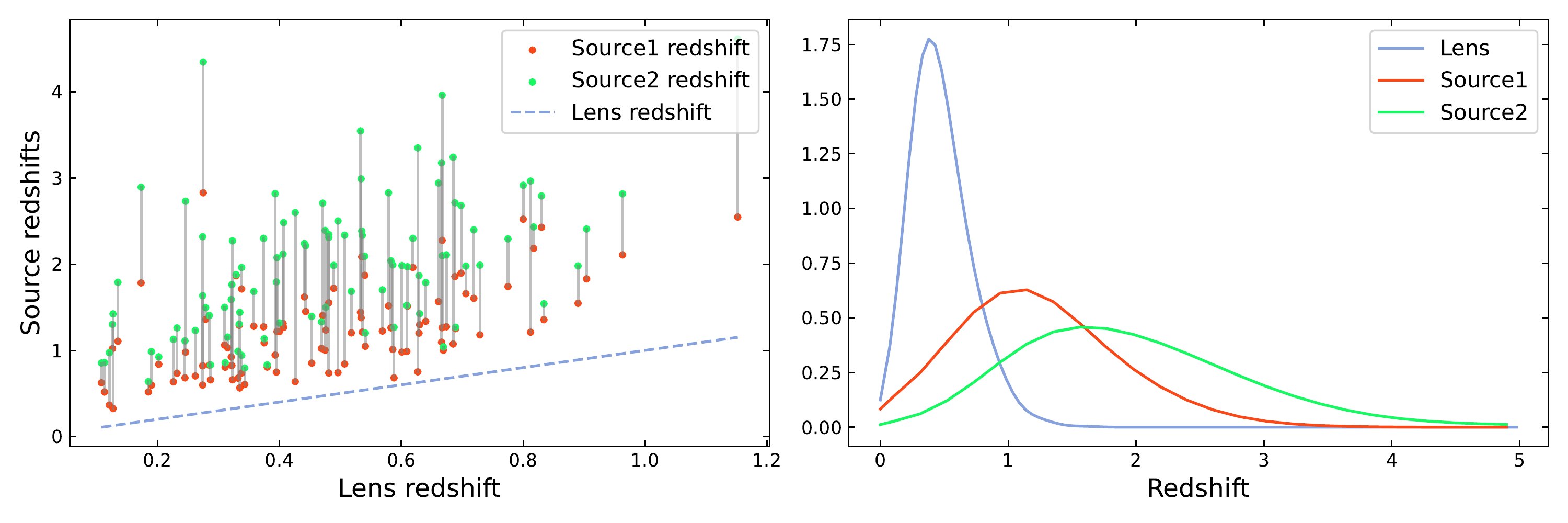}
    \caption{\textit{Left}: the redshift distribution of 100 example DSPLs from {\tt LensPop}. \textit{Right}: The histogram represents the redshift distribution of 1\,700 systems drawn from {\tt LensPop}. The blue line indicates the lens redshift, while the red and green dashed lines mark the redshifts of source 1 and source 2, respectively. }
    \label{fig:redshift_distribution}
\end{figure*}

\begin{figure*}
    \centering
    \includegraphics[width=1\textwidth]{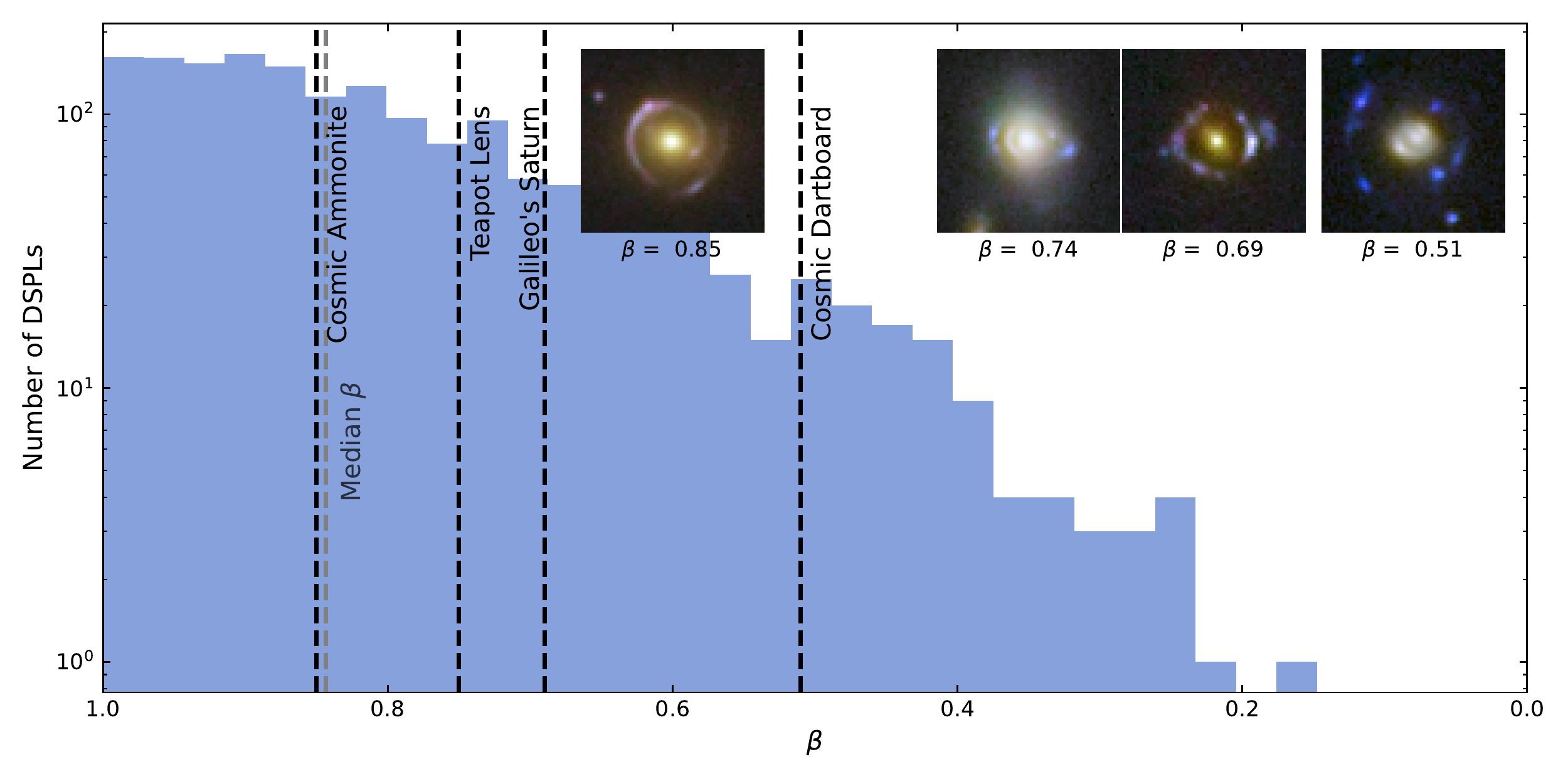}
    \caption{Distribution of $\beta$ from the forecast. The black dashed lines mark the measured $\beta$ values for DSPLs discovered in Q1, and the gray dashed line indicates the median $\beta$ value in the forecast catalogue. }
    \label{fig:beta_distribution}
\end{figure*}

This lens was first discovered during visual inspection of high-velocity dispersion galaxies from Dark Energy Spectroscopic Instrument \citep{Q1-SP052} as part of the effort to build a training set for machine learning. It was later rediscovered in Space Warps and Galaxy Judges, ranking 622 in the {\tt Zoobot} and 237 in Galaxy Judges. It was classified as a 'Grade B' lens.
As shown in Fig. \ref{fig:crackpot}, the system features two blue arcs at different radii. It is a lens with two bright lensing elliptical galaxies. While the counter-image of the top arc is clearly visible at the bottom left of the lens, the counter-image of the inner arc is only faintly visible in the lens light-subtracted image, making its identification less convincing.

Figure \ref{fig:EDFN10} shows the lens model of the Crackpot lens. Assuming the faint inner arc is real, the double-source-plane lens model suggests that the two images have similar Einstein radii. However, the presence of a nearby galaxy at the bottom of the image complicates the mass profile. Based on these factors, we believe that this lens is unlikely to be a DSPL, but rather a lens with two sources at the same redshift. We have named this system the Crackpot as a tongue-in-cheek reference to the fact it is not a Jackpot lens after all.





\section{Expected rate and population of DSPLs in \Euclid}
\label{sec:catalog}
The \Euclid survey should enable the construction of a sample of approximately 1700 galaxy-scale DSPL systems. This forecast is derived from the {\tt LensPop} package \citep{Collett:2015roa}, modified to include multiple background sources (we neglect the mass of the first source). We also impose more stringent constraints than the {\tt LensPop} defaults. Specifically, we require that both sources have one or more arcs of length $\ang{;;0.3}$ -- this ensures a reasonable possibility that the density slope of the lens can be recovered from \Euclid imaging alone. The redshift distribution for lenses and sources is shown in Fig. \ref{fig:redshift_distribution}. Although most compound lenses are at $z \approx 0.5$, a significant fraction of lenses are at $z<0.2$ and at $z>1$, which is promising for a precise constraint on the mass distribution in lenses. Of the 1700 expected \Euclid DSPLs, approximately $6 \pm 3$ should fall inside the Q1 footprint (assuming that the DSPLs are randomly distributed). The entire \Euclid survey will cover about \num{14000}\,deg$^2$, whereas the Q1 footprint considered here spans only 63\,deg$^2$. The forecasted number of systems in such a small area is expected to be modest, so even a difference of a few lenses (e.g., finding four instead of the predicted six) can be attributed to small-number statistics. Poisson fluctuations and cosmic variance can easily shift the observed number by a few systems relative to theoretical forecasts.

Our four DSPLs already represent a more than doubling of the galaxy-galaxy DSPL population. They should be useful for cosmography, but without any redshift information, it is impossible to map $\beta$ values onto cosmological parameter constraints.
Notably, the \cdl ~has $\beta \approx 0.51$ which is quite rare based on the $\beta$ distribution of the mock catalogue as shown in Fig. \ref{fig:beta_distribution} (only 6$\%$ of the DSPLs have $\beta$ < 0.5, so the probability of having at least one DSPL with $\beta$ < 0.5 in four systems is 0.2). This indicates that our forecast might underestimate the DSPL population observed by \Euclid. Theoretically, smaller $\beta$ values (two source galaxies at very different redshift) are more sensitive to the cosmology parameters, since the DSPL with a decreased $\beta$ provides enhanced geometric leverage and mitigates degeneracies amongst cosmological parameters. For instance, when $\beta \approx 1$ ($z_\mathrm{s1} \approx z_\mathrm{s2}$), variations in cosmological parameters can result in $\beta$ remaining close 1.

\section{Conclusions}
\label{sec:conclusion}
We have identified four new galaxy-scale DSPL candidates in Q1, significantly extending the limited sample of known DSPL systems. Through detailed lens modelling, we strengthened the plausibility of their DSPL nature and derived the $\beta$ (cosmological scaling factor which represents a distance ratio) parameters for each candidate, demonstrating their potential ability in cosmological studies. We also have a handful of tentative DSPL candidates for which higher-resolution imaging and spectroscopic redshifts are needed to make definitive conclusions.

We have piggybacked on the single source plane lens search to find interesting DSPL candidates. We did not train any machine or human classifiers to explicitly look for DSPLs. One consequence of not creating ML lens-finders trained on DSPLs is that they do not learn to value DSPL systems higher than regular lenses. This means the DSPLs will not necessarily be ranked exceptionally highly amongst the lenses, which was what was found in Q1 \citep{Q1-SP053}. Future \Euclid single-plane-lens searches will be over larger areas, meaning we will be unable to visually inspect the same proportion of the total sample as we have done in Q1: scaling up the proportion of the total number of images visually inspected in Q1 to the full \Euclid survey corresponds to visually inspecting 6 million images, which is likely intractable. Both the DSPLs and the single-source-plane lenses found in Q1 will likely enable significant improvement in the ML performance, and training ML models on DSPLs should increase the probability of these being recovered in future \Euclid lens searches. However, it may be possible that a dedicated DSPL search will be needed for \Euclid DR1 and beyond.

It is possible that some DSPLs remain to be found in Q1, either because they were ranked poorly by the single source plane lens ML classifiers or because they have been missed in visual inspection (we do not yet know how our methodology performs with small Einstein radius systems). We have also missed any lenses that do not pass our initial lens cuts ($I_\mathrm{E}$ < 22.5 mag). Some lenses are expected to be faint in \IE \citep[\IE > 24 mag,][]{Collett:2015roa} although these would typically be low mass or high redshift lenses (e.g., lens redshift larger than 1.5). Either scenario is less likely to produce a \Euclid-detectable DSPL than a galaxy that is more luminous in \IE. Our forecasts indicate that the full \Euclid survey at the end of operations is expected to uncover 1700 DSPLs, whilst extrapolating up from our four DSPLs in 63 deg$^2$ yields 1000 ($\pm$ 500) DSPLs in the full survey. Whether we ultimately find 1000 or 1700 lenses in the full \Euclid data set, it is already clear that \Euclid will revolutionise research using DSPLs.

\begin{acknowledgements}
Numerical computations were done on the Sciama High Performance Compute (HPC) cluster, which is supported by the ICG, SEPNet, and the University of Portsmouth.

This work has received funding from the European Research Council (ERC) under the European Union's Horizon 2020 research and innovation program (LensEra: grant agreement No 945536). TC is funded by the Royal Society through a University Research Fellowship. CT acknowledges the INAF grant 2022 LEMON.
For the purpose of open access, the authors have applied a Creative Commons Attribution (CC BY) license to any Author Accepted Manuscript version arising.

\AckEC
\AckQone
\end{acknowledgements}

\section{Data Availability}
The Q1 data is available at the \Euclid science archive. The forecast population of DSPLs are avaliavle at {\tt LensPop} github repository. The derived data products are available upon request from the corresponding author.
%
%

\bibliography{mycitation, Euclid, Q1}

%

\begin{appendix}
  \onecolumn 

\section{Other DSPL lens candidates}
\label{sec:imposter}

\begin{figure*}
    \centering
    \includegraphics[width=0.80\textwidth]{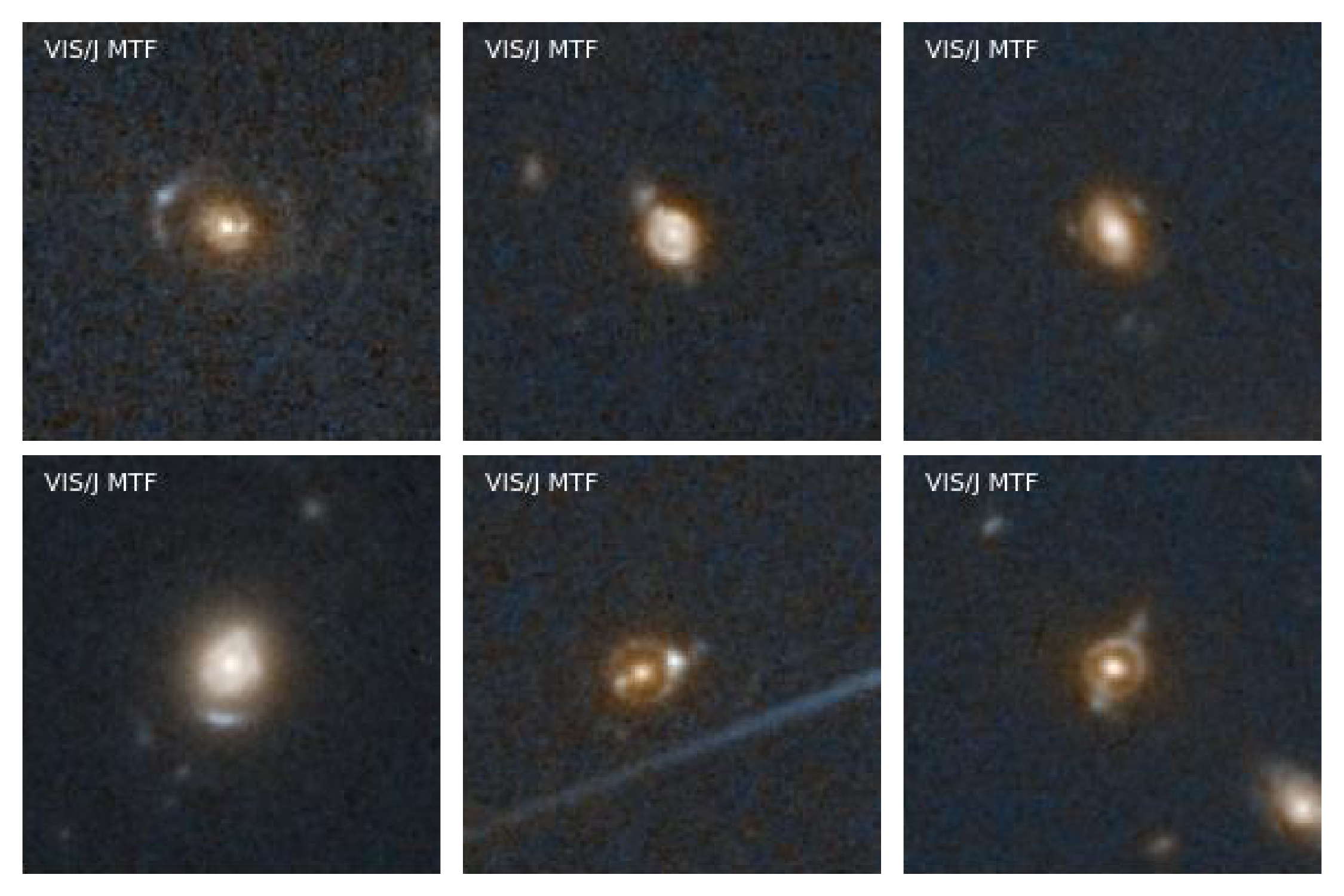}
    \caption{Low probability DSPL candidates selected among the top 1000 ranked galaxies in Galaxy Judges.}
    \label{fig:imposters_good}
\end{figure*}

\begin{figure*}
    \centering
    \includegraphics[width=0.82\textwidth]{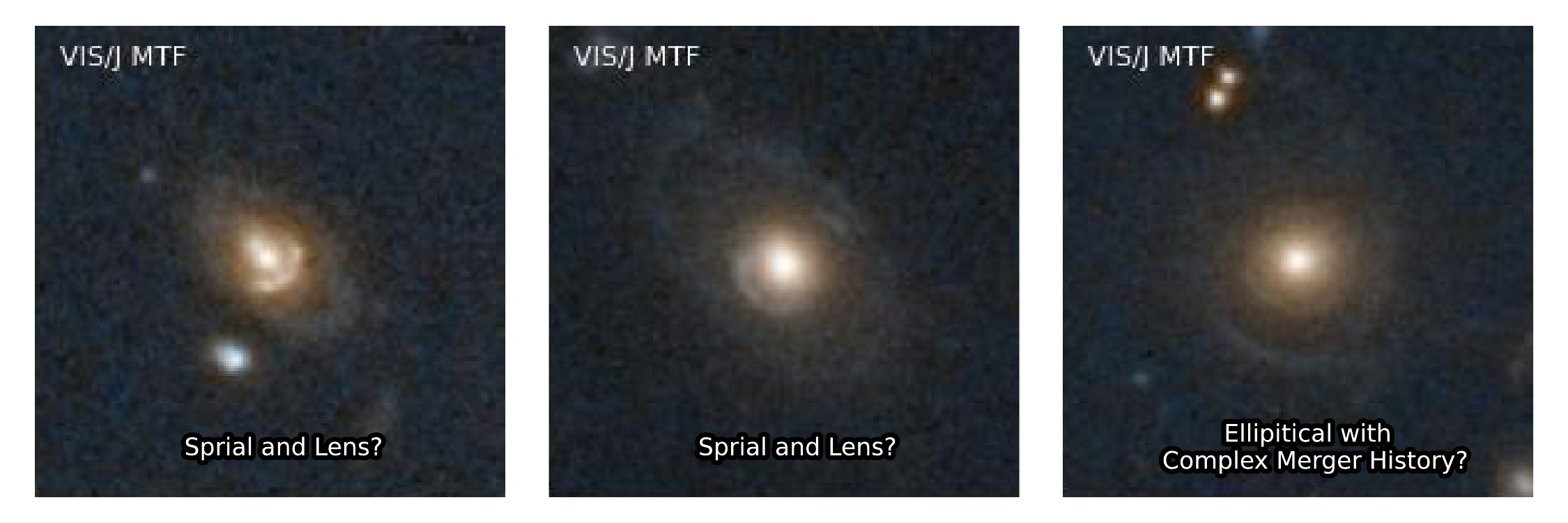}
    \caption{Candidate DSPL imposters selected among the top 1000 ranked galaxies in Galaxy Judges.}
    \label{fig:imposters_bad}
\end{figure*}

Due to the large survey area and high-resolution images provided by \Euclid, rarer single-plane lensing configurations, and false positives can be discovered which could be misidentified as possible DSPL systems. Figure \ref{fig:imposters_good} shows a sample of lens candidates in Q1 which might be DSPLs but we can not classify them with confidence due to their small Einstein radii. Figure \ref{fig:imposters_bad} shows a sample of images that could be confused for DSPLs but we believe are most likely contaminants. Here we outline several candidate configurations that create images that are similar to DSPLs:

\begin{itemize}
    \item Double cusp configuration: When a source is positioned near a cusp of the tangential caustic, three closely spaced lensed images appear in the lens plane, while a fourth, more isolated and typically less magnified image, forms at a closer distance. If two sources within the same source plane are located near two opposing cusps of the diamond-shaped caustic, the resulting lensing configuration can produce a ring-like structure near the lens, along with two smaller arcs inside the ring. This configuration may resemble a DSPL system and could be mistakenly identified as such. 
    
    \item Spiral galaxy and lens: In strong lensing systems where the deflector is a spiral galaxy, features like spiral arms or star-forming rings can sometimes be misidentified as secondary rings at different (smaller or larger) radii. This is because star-forming regions often exhibit a distinct blue colour, similar to that of background lensed sources. Such cases are not common, as face-on spiral galaxies typically have low projected surface mass density, resulting in a small lensing cross-section. However, at least two possible cases were found in our sample although they may not even be lenses. This type of lens has its unique value but it's beyond the scope of this paper.

    \item Spiral galaxy with multiple arms: As described above, spiral arms can resemble lensed arcs, making certain spiral galaxies appear similar to DSPL systems. This is particularly true when the central bulge of the galaxy is large enough to mimic the appearance of an elliptical galaxy. However, such systems can typically be easily ruled out through lens modelling. 

    \item Elliptical lens with complex shells: In addition to spiral arms, the shell structures of elliptical galaxies can also resemble lensed arcs. These features become more apparent after subtracting the lens light in preliminary lens modelling \citep{Q1-SP048}. However, such structures are often noisy, as they are typically buried beneath the bright lens light, making them more challenging to identify without careful analysis.

    \item Lens system with nearby galaxies: galaxies located close in projection to a lens but outside the multiply imaged region can mimic an image from a doubly lensed system. An expert might incorrectly guess that there is an undetected counter-image of this galaxy close to the centre of the primary lens.

\end{itemize}
The scenarios described above represent the most common cases where DSPL systems can be misidentified. Conversely, a genuine DSPL system can also be misclassified as one of these cases. To make a definitive determination either lens modelling of higher angular resolution images or spectroscopic redshifts for each component, are required.
\end{appendix}

\end{document}